%% file: main.tex
\documentclass[runningheads]{llncs}
\usepackage{color}
\usepackage{amsmath}
\usepackage{amsfonts}
\usepackage{hyperref}
\usepackage{proof}
\usepackage{bussproofs}
\usepackage{stmaryrd}

\usepackage{amsthm}
\usepackage{tikz}
\usetikzlibrary{calc,arrows,intersections}
\usetikzlibrary{shapes,fit}
\usetikzlibrary{positioning}
\usetikzlibrary{patterns}
\usepackage{subcaption}
\usepackage{sansmath}
\usepackage[inline]{enumitem}

\input{macros}

\usepackage{xcolor}
\definecolor[named]{ACMBlue}{cmyk}{1,0.1,0,0.1}
\definecolor[named]{ACMYellow}{cmyk}{0,0.16,1,0}
\definecolor[named]{ACMOrange}{cmyk}{0,0.42,1,0.01}
\definecolor[named]{ACMRed}{cmyk}{0,0.90,0.86,0}
\definecolor[named]{ACMLightBlue}{cmyk}{0.49,0.01,0,0}
\definecolor[named]{ACMGreen}{cmyk}{0.20,0,1,0.19}
\definecolor[named]{ACMPurple}{cmyk}{0.55,1,0,0.15}
\definecolor[named]{ACMDarkBlue}{cmyk}{1,0.58,0,0.21}

\newcommand{\RAPHAELLE}[1]{#1}

\begin{document}
\title{On the Versatility of Open Logical Relations\thanks{The Second and Fourth Authors are 
supported by the ANR project 16CE250011 REPAS, the ERC Consolidator Grant 
DIAPASoN -- DLV-818616, and the MIUR PRIN 201784YSZ5 ASPRA.
}}
\subtitle{Continuity, Automatic Differentiation,\\ and a Containment Theorem}
\titlerunning{On the Versatility of Logical Relations}
\author{Gilles Barthe\inst{1,4}
  \and
Rapha\"elle Crubill\'e\inst{4}
  \and
Ugo Dal Lago\inst{2,3}
  \and
Francesco Gavazzo\inst{2,3,4}
}
\authorrunning{G. Barthe et al.}
%
\institute{
  MPI for Security and Privacy, Germany
  \and
  University of Bologna, Italy
  \and
  INRIA Sophia Antipolis, France
  \and
  IMDEA Software Institute, Spain
}
\maketitle              
\begin{abstract}
  Logical relations are one of the most powerful techniques in the theory
  of programming languages, and have been used extensively for proving
  properties of a variety of higher-order calculi.  However, there are
  properties that cannot be immediately proved by means of logical
  relations, for instance program continuity and differentiability in
  higher-order languages extended with real-valued functions.
  Informally, the problem stems from the fact that these properties
  are naturally expressed on terms of non-ground type (or, equivalently,
  on open terms of base type), and there is no apparent good definition
  for a base case (i.e. for closed terms of ground types). To overcome this
  issue, we study a generalization of the concept of a logical
  relation, called \emph{open logical relation}, and prove that
  it can be fruitfully applied in several contexts in which the
  property of interest is about expressions of first-order type. Our
  setting is a simply-typed $\lambda$-calculus enriched with real
  numbers and real-valued first-order functions from a given set, 
  such as the one of continuous or differentiable functions. We first
  prove a containment theorem stating that for any such a collection
  of functions including projection functions and closed under function composition, 
  any well-typed term of first-order type denotes a
  function belonging to that collection.  Then, we show by way of open
  logical relations the correctness of the core of a recently
  published algorithm for forward automatic differentiation. Finally,
  we define a refinement-based type system for local continuity in an
  extension of our calculus with conditionals, and prove the soundness
  of the type system using open logical relations.

\end{abstract}

\section{Introduction}
\label{sect:introduction}
Logical relations have been extremely successful as a way of proving
equivalence between concrete programs as well as correctness of
program transformations. In their ``unary'' version, they also are a
formidable tool to prove termination of typable programs, through the
so-called \emph{reducibility} technique. The class of programming
languages in which these techniques have been instantiated includes
not only higher-order calculi with simple types, but also calculi with
recursion \cite{DBLP:journals/toplas/AppelM01,DBLP:conf/esop/Ahmed06,DBLP:journals/entcs/CraryH07}, 
various kinds of effects \cite{DBLP:conf/fossacs/BizjakB15,DBLP:journals/iandc/BirkedalJST16,DBLP:journals/jfp/DreyerNB12,DBLP:conf/birthday/Hofmann15,DBLP:conf/popl/Benton0N14,DBLP:journals/pacmpl/BiernackiPPS18,Goubault}, 
and concurrency \cite{DBLP:conf/popl/TuronTABD13,DBLP:conf/csl/BirkedalST12}.

Without any aim to be precise, let us see how reducibility works, in
the setting of a simply typed calculus. The main idea is to define, by
induction on the structure of types, the concept of a well-behaved
program, where in the base case one simply makes reference to the
underlying notion of observation (e.g. being strong normalizing),
while the more interesting case is handled by stipulating that
reducible higher-order terms are those which maps reducible terms to
reducible terms, this way exploiting the inductive nature of simple
types. One can even go beyond the basic setting of simple types, and
extend reducibility to, e.g., languages with recursive types 
\cite{DBLP:journals/entcs/CraryH07,DBLP:conf/esop/Ahmed06} or
even untyped languages \cite{DBLP:conf/esop/OwensMKT16} by means of techniques such as
step-indexing \cite{DBLP:journals/toplas/AppelM01}.

The same kind of recipe works in a relational setting, where one wants
to \emph{compare} programs rather than merely \emph{proving
  properties} about them. Again, two terms are equivalent at base
types if they have the same observable behaviour, while at higher
types one wants that equivalent terms are those which maps equivalent
arguments to equivalent results.


There are cases, however, in which the property one observes, or the
property in which the underlying notion of program equivalence or
correctness is based, is formulated for types which are \emph{not}
ground (or equivalently, it is formulated for open expressions). As an
example, one could be interested in proving that in a higher-order
type system all \emph{first-order} expressions compute numerical
functions of a specific kind, for example, continuous or derivable
ones. We call such properties \emph{first-order
  properties}\footnote{To avoid misunderstandings, we emphasize that
  we use first-order properties to refer to properties of expressions
  of first-order types---and not in relation with definability of
  properties in first-order predicate logic.}.  As we will describe in
Section~\ref{sect:gap} below, logical relations do not seem to be
applicable \emph{off-the-shelf} to these cases. Informally, this is
due to the fact that we cannot start by defining a base case for
ground types and then build the relation inductively.

In this paper, we show that
logical relations and reducibility
can deal with first-order properties in a compositional way
without altering their nature. 
The main idea behind the
resulting definition, known as \emph{open logical relations} 
\cite{DBLP:conf/aplas/ZhaoZZ10},
consists in parameterizing 
the set of related terms
of a certain type (or the underlying reducibility set)
on a \emph{ground environment}, this way turning it into a
set of pairs of \emph{open terms}. As a consequence, one can define the
target first-order property in a natural way.

Generalizations of logical relations to open terms 
have been used by several 
authors, and in several (oftentimes unrelated) 
contexts (see, for instance, 
\cite{DBLP:conf/icfp/BowmanA15,DBLP:conf/tlca/JungT93,DBLP:conf/mfcs/PittsS93,DBLP:conf/ppdp/Fiore02,DBLP:conf/lics/StatonYWHK16}). 
In this paper, we show how open logical relations 
constitute a powerful technique to systematically prove 
first-order properties of programs. In this respect,
the paper technical contributions are applications 
of open logical relations to three distinct problems.
\begin{varitemize}
\item In Section~\ref{section:conservation-theorem}, we use open
  logical relations to prove a general \ctheorem. Such a theorem
  serves as a vehicle to introduce open logical relations but is also of
  independent interest. 
  The theorem states
  that given a collection $\funsymbols$ of real-valued functions 
  including projections and closed under function composition, any
  first-order term of a simply-typed $\lambda$-calculus endowed with
  operators computing functions in $\funsymbols$, computes a function
  in $\funsymbols$. 
  As an instance of such a result, we
  see that any first-order term in a simply-typed $\lambda$-calculus
  extended with primitives for continuous functions, computes a
  continuous function. 
  Although the \ctheorem\ can be derived from previous results by Lafont 
  \cite{lafont1988logiques} 
  (see Section~\ref{sect:related-work}), our proof is purely syntactical 
  and consists of a
  straightforward application of open logical relation. 
\item In Section~\ref{sect:automatic-differentiation}, we use open
  logical relations to prove correctness of a core algorithm for
  forward automatic differentiation of simply-typed terms. The
  algorithm is a fragment of \cite{Efficient/AD/jones}.  More
  specifically, any first-order term is proved to be mapped to another
  first-order term computing its derivative, in the usual sense of
  mathematical analysis.  This goes beyond the \ctheorem\ by dealing
  with relational properties. 
  
\item In Section~\ref{sect:continuity}, we consider an extended language with
  an if-then-else construction. When dealing with continuity, 
  the introduction of conditionals
  invalidates the \ctheorem, since conditionals naturally introduce
  discontinuities. To overcome this deficiency, we introduce a refinement type system 
  ensuring that
  first-order typable terms are continuous functions on some intended
  domain, and use open logical relations to prove the soundness of the
  type system. 
\end{varitemize}

Due to space constraints, many details have to be omitted, 
but can be found in an Extended Version of this work~\cite{EV}.

\section{The Playground}\label{sect:the_playground}

In order to facilitate the communication of the main ideas behind 
open logical relations and their applications, 
this paper deals with several vehicle calculi. All such calculi can be seen 
as derived from a unique calculus, denoted by $\STLambdareal$, which thus provides the 
common ground for our inquiry. 
The calculus $\STLambdareal$ is obtained by adding to the simply typed 
$\lambda$-calculus with product and arrow types (which we denote by 
$\STLambda$) a ground type $\typereal$ for real numbers and constants 
$\makereal{r}$ of type $\typereal$, for each real number $r$.

Given a collection $\funsymbols$ of real-valued functions, i.e. 
functions $f: \mathbb{R}^n \to \mathbb{R}$ (with $n \geq 1$), 
we endow $\STLambdareal$ with an operator $\makereal{\funsymbol}$, 
for any $\funsymbol \in \funsymbols$,
whose intended meaning 
is that whenever $\termone_1, \hh, \termone_n$ represent real numbers 
$r_1, \hh, r_n$, then 
$\makereal{\funsymbol}(\termone_1, \hh, \termone_n)$ represents 
$\funsymbol(r_1, \hh, r_n)$. We call the resulting calculus 
$\STLambdarealfun$. 
Depending on the application we are interested in, 
we will take as $\funsymbols$ specific
collections of real-valued functions, 
such as continuous or differentiable functions. 

The syntax and static semantics of $\STLambdarealfun$ are defined in 
\autoref{figure:static-semantics-lambda}, where 
$\funsymbol: \mathbb{R}^n \to \mathbb{R}$ belongs to 
$\funsymbols$.
The statics of $\STLambdarealfun$ is based on judgments of the form 
$\envone \imp \termone: \typeone$, which have the usual intended meaning. 
We adopt standard syntactic conventions as in 
\cite{barendregt_lambda_1984}, notably the so-called variable convention. 
In particular, we denote by $FV(\termone)$ the collection 
of free variables of $\termone$ and by $\subst{\termtwo}{\varone}{\termone}$ 
the capture-avoiding substitution of 
the expression $\termone$ for all free occurrences of $\varone$ in $\termtwo$.
\begin{center}
\begin{figure}[htbp]
  \fbox{
    \begin{minipage}{.97\textwidth}
{\center
      \begin{align*}
\typeone &= \typereal 
\mid \typeone \times \typeone 
\mid \typeone \to \typeone
&
\envone &::= \cdot \mid \varone: \typeone, \envone
\end{align*}
\vspace{-0.7cm}
\begin{align*}
\termone &::= \varone 
\mid \makereal{r}
\mid \makereal{\funsymbol}(\termone, \hh, \termone)
\mid \abs{\varone}{\termone} 
\mid \termone \termone 
\mid (\termone, \termone) 
\mid \fst{\termone} 
\mid \snd{\termone}
\end{align*}
\[
\infer{\envone, \varone: \typeone \imp \varone: \typeone}{\phantom{\varone}}
\qquad
\infer{\envone \imp \makereal{r}: \typereal}{\phantom{\varone}}
\qquad
\infer
{\envone \imp \makereal{\funsymbol}(\termone_1, \hh, \termone_n): \typereal}
{\envone \imp \termone_1: \typereal 
& \cc
& \envone \imp \termone_n: \typereal
}
\qquad
\infer
{\envone \imp \abs{\varone}{\termone}: \typeone_1 \to \typeone_2}
{\envone, \varone:\typeone_1 \imp \termone: \typeone_2}
\]
$\vspace{-0.3cm}$
\[
\infer
{\envone \imp \termtwo \termone: \typeone_2}
{\envone \imp \termtwo : \typeone_1 \to \typeone_2
&
\envone \imp \termone: \typeone_1}
\qquad
\infer
{\envone \imp (\termone_1, \termone_2): \typeone \times \typetwo}
{\envone \imp \termone_1: \typeone
& \envone \imp \termone_2: \typetwo}
\qquad 
\infer[(i \in \{1,2\})]
{\envone \imp \ith{\termone}: \typeone_i}
{\envone \imp \termone: \typeone_1 \times \typeone_2}
\]
}
\end{minipage}}
\caption{Static semantics of $\STLambdarealfun$.}
\label{figure:static-semantics-lambda}
\end{figure}
\end{center}
We do not confine ourselves with a fixed operational semantics 
(e.g. with a call-by-value operational semantics), but take advantage 
of the simply-typed nature of $\STLambdarealfun$ and opt for 
a set-theoretic denotational semantics. The category of sets and 
functions being cartesian closed, the denotational semantics of 
$\STLambdarealfun$ is standard and associates to any judgment
$x_1: \typeone_1, \hh, x_n: \typeone_n \imp \termone: \typeone$, 
a function $\sem{x_1: \typeone_1, \hh, x_n: \typeone_n \imp \termone: \typeone}: 
\prod_i \sem{\typeone_i} \to \sem{\typeone}$, where $\sem{\typeone}$---the 
\emph{semantics} of $\typeone$---is thus defined:
\begin{align*}
\sem{\typereal} &= \mathbb{R};
&
\sem{\typeone_1 \to \typeone_2} &= \sem{\typeone_2}^{\sem{\typeone_1}};
& 
\sem{\typeone_1 \times \typeone_2} &= \sem{\typeone_1} \times \sem{\typeone_2}.
\end{align*}
Due to space constraints, we omit the definition of $\sem{\envone \imp \termone: \typeone}$ and refer the reader to any textbook on the subject 
(such as \cite{DBLP:books/daglib/0085577}).

\section{A Fundamental Gap}\label{sect:gap}

In this section, we will look informally at a problem which, apparently, cannot
be solved using reducibility or logical relations. This serves both as
a motivating example and as a justification of some of the design
choices we had to do when designing open logical relations.

Consider the simply-typed $\lambda$-calculus $\STLambda$, the
prototypical example of a well-behaved higher-order functional
programming language. As is well known, $\STLambda$ is strongly
normalizing and the technique of logical relations can be applied
on-the-nose. The proof of strong normalization for $\STLambda$ is
structured around the definition of a family of reducibility sets of
\emph{closed} terms $\{\mathit{Red}_{\typeone}\}_{\typeone}$, indexed
by types.  At any atomic type $\typeone$, $\mathit{Red}_{\typeone}$ is
defined as the set of terms (of type $\typeone$) 
having the property of interest, i.e. as the collection of strongly normalizing 
terms. 
The set $\mathit{Red}_{\typeone_1\to\typeone_2}$,
instead, contains those terms which, when applied to a term in
$\mathit{Red}_{\typeone_1}$, returns a term in
$\mathit{Red}_{\typeone_2}$. Reducibility sets are afterwards
generalised to open terms, and finally all typable terms are shown to
be reducible.

Let us now consider the calculus $\STLambdarealfun$,
where $\funsymbols$ contains the addition and multiplication functions 
only.
This language has already been
considered in the literature, under the name of \emph{higher-order
  polynomials}~\cite{DBLP:conf/focs/CookK89,DBLP:journals/siamcomp/KapronC96}, which are crucial tools in higher-order
complexity theory and resource analysis.  Now, let us ask ourselves
the following question: can we say anything about the nature of those
functions $\RR^n\to\RR$ which are denoted by (closed) terms of type
$\typereal^n\to\typereal$? Of course, all the polynomials on the real
field can be represented, but can we go beyond, thanks to higher-order
constructions? The answer is negative: terms of type
$\typereal^n \to\typereal$ represent all \emph{and only} the
polynomials~\cite{baillot2012higher,DBLP:journals/corr/abs-1005-0524}. This result is an instance of the general
containment theorem mentioned at the end of
Section~\ref{sect:introduction}.

Let us now focus on proofs of this containment result. It turns out
that proofs from the literature are not compositional, and rely
on\lq\lq heavyweight\rq\rq\ tools, including strong normalization of
$\STLambda$ and soundness of the underlying operational semantics.  In
fact, it is not immediate to prove the result using
usual reducibility arguments, precisely because there is no obvious
choice for the base case.  If, for example, we define
$\mathit{Red}_{\typereal}$ as the set of terms strongly normalizing to
a numeral, $\mathit{Red}_{\typereal^n\to\typereal}$ as the set of
polynomials, and for any other type as usual, we soon get into
troubles: indeed, we would like the two sets of functions
$$
\mathit{Red}_{\typereal\times\typereal\to\typereal}
\qquad\qquad
\mathit{Red}_{\typereal\to(\typereal\to\typereal)}
$$
to denote \emph{essentially} the same set of functions, modulo the
adjoint between $\RR^2\to\RR$ and $\RR\to(\RR\to\RR)$. But this is
clearly not the case: just consider the function $\funsymbol$ in
$\RR\to(\RR\to\RR)$
$$
\funsymbol(x)=
\left\{
\begin{array}{ll}
  \lambda y.y & \mbox{if $x\geq 0$}\\
  \lambda y.y+1 & \mbox{if $x<0$}
\end{array}
\right.
$$
Clearly, $\funsymbol$ turns any \emph{fixed} real number to a polynomial,
but when curried, it is far from being a polynomial. In other
words, reducibility seems apparently inadequate to capture situations like
the one above, in which the ``base case'' is not the one of ground type,
but rather the one of first-order types.

Before proceeding any further, it is useful to fix the boundaries of
our investigation. We are interested in proving that (the semantics
of) programs of first-order type $\typereal^n \to \typereal$ enjoy
first-order properties, such as continuity or differentiability, under
their standard interpretation in calculus and real analysis. More
specifically, our results do not cover notions of continuity and
differentiability studied in fields such as (exact) real-number
computation \cite{Vuillemin:1990:ERC:101793.101805} 
or computable analysis \cite{weihrauch2000computable}, which have a strong
domain-theoretical flavor, and higher-order generalizations of
continuity and differentiability (see, e.g.,
\cite{DBLP:journals/entcs/Edalat00,DBLP:conf/lics/EdalatL02,DBLP:conf/fossacs/GianantonioE13,DBLP:journals/iandc/EscardoH09}). We
leave for future work the study of open logical relations in these
settings. What this paper aims to provide, is a family of \emph{lightweight}
techniques that can be used to show that practical properties of
interest of real-valued functions are guaranteed to hold when programs
are written taking advantage of higher-order constructors.  We believe
that the three case studies we present in this paper are both a way to point to
the practical scenarios we have in mind and of witnessing the versatility of
our methodology.

\section{Warming Up: A Containment Theorem}
\label{section:conservation-theorem}

In this section we introduce open logical relations in their unary version 
(i.e. open logical predicates). We do so by proving the 
following Containment Theorem. 

\begin{theorem}[\ctheorem]
\label{thm:conservation-theorem}
Let $\funsymbols$ be a collection of real-valued functions including 
projections and closed under function composition. Then, any 
$\STLambdarealfun$ term 
$\varone_1: \typereal, \hh, \varone_n: \typereal_n \imp \termone: \typereal$ 
denotes a function (from $\mathbb{R}^n$ to $\mathbb{R}$) in 
$\funsymbols$. That is, $\sem{\varone_1: \typereal, \hh, \varone_n: \typereal_n \imp \termone: \typereal} \in \funsymbols$.
\end{theorem}

As already remarked in previous sections, notable instances of 
Theorem~\ref{thm:conservation-theorem} are obtained by taking 
$\funsymbols$ as the collection of continuous functions, 
or as the collection of polynomials. 

Our strategy to prove Theorem~\ref{thm:conservation-theorem} 
consists in defining a logical predicate, 
denoted by $\funpred$, ensuring the denotation of programs of a first-order type to 
be in $\funsymbols$, and hereditary preserving 
this property at higher-order types. 
However, $\funsymbols$ being a property of real-valued 
functions---and the denotation of an \emph{open} term of the form 
$\varone_1: \typereal, \hh, \varone_1: \typereal \imp \termone: \typereal$ being 
such a function---we shall work with open terms with free 
variables of type $\typereal$ and parametrize the candidate logical 
predicate by types \emph{and} environments $\envreal$ containing such variables. 

This way, we obtain a family of logical predicates 
$\funpred_{\typeone}^{\envreal}$ acting on terms of the form 
$\envreal \imp \termone: \typeone$. As a consequence, when considering the ground type 
$\typereal$ and an environment $\envreal = \varone_1: \typereal, \hh, \varone_n: \typereal$,
 we obtain a predicate $\funpred_{\typereal}^{\envreal}$ on expressions 
$\envreal \imp \termone: \typereal$ which naturally corresponds to functions 
from $\mathbb{R}^n$ to $\mathbb{R}$, for which belonging to $\funsymbols$ is indeed 
meaningful.

\begin{definition}[Open Logical Predicate]
\label{def:open-logical-predicate} 
Let $\envreal = \varone_1: \typereal, \hh, 
\varone_n: \typereal$ be a fixed environment.
W define the type-indexed family of predicates
$\funpred^{\envreal}_{\typeone}$ by induction 
on $\typeone$ as follows:
\begin{align*}
  \termone \in \funpred^{\envreal}_{\typereal} 
  &\iff 
  (
    \envreal \imp \termone: \typereal \wedge
    \sem{\envreal \imp \termone: \typereal} \in \funsymbols
  )
  \\
  \termone \in \funpred^{\envreal}_{\typeone_1 \to \typeone_2} 
  &\iff
   (
    \envreal \imp \termone: \typeone_1 \to \typeone_2
    \wedge
    \forall \termtwo \in \funpred^{\envreal}_{\typeone_1}.\ 
    \termone\termtwo \in \funpred^{\envreal}_{\typeone_2}
  )
  \\
  \termone \in \funpred^{\envreal}_{\typeone_1 \times \typeone_2} 
  &\iff
   (
    \envreal \imp \termone: \typeone_1 \times \typeone_2 
    \wedge 
    \forall i \in \{1,2\}.\ 
    \ith{\termone} \in \funpred^{\envreal}_{\typeone_i}
    ).
\end{align*}
We extend $\funpred^{\envreal}_{\typeone}$ to the predicate 
$\funpred^{\envone,\envreal}_{\typeone}$, 
where $\envone$ ranges over arbitrary environments (possibly containing 
variables of type $\typereal$) as follows: 
\begin{align*}
\termone \in \funpred^{\envone, \envreal}_{\typeone} 
&\iff 
(
\envone, \envreal \imp \termone: \typeone 
\wedge
\forall \substone.\ 
\substone \in \funpred^{\envone}_{\envreal} \implies 
\termone \substone \in \funpred^{\envreal}_{\typeone} 
).
\end{align*}
Here, $\substone$ ranges over substitutions\footnote{We write 
$\termone \substone$ for the result of applying $\substone$ to 
variables in $\termone$.} and $
\substone \in \funpred^{\envone}_{\envreal} 
$ holds if the support of $\substone$ is $\envone$ and 
$\substone(\varone) \in \funpred^{\envreal}_{\typeone}$, 
for any $(\varone: \typeone) \in \envone$.
\end{definition}

Notice that Definition~\ref{def:open-logical-predicate} 
ensures first-order real-valued functions to be in $\funsymbols$, 
and asks for such a property to be hereditary preserved at higher-order 
types. Lemma~\ref{lemma:fundamental-lemma-logical-predicate} states 
that these conditions are indeed sufficient to guarantee any $\STLambdarealfun$ 
term 
$\envreal \imp \termone: \typereal$ 
to denote a function in $\funsymbols$.

\begin{lemma}[Fundamental Lemma]
\label{lemma:fundamental-lemma-logical-predicate}
For all environments $\envone, \envreal$ as above, and 
for any expression $\envone, \envreal \imp \termone: \typeone$, 
we have $\termone \in \funpred^{\envone, \envreal}_{\typeone}$. 
\end{lemma}

\begin{proof}
By induction on $\termone$, observing that
$\funpred_{\typeone}^{\envreal}$ is closed under denotational 
semantics: if $\termtwo \in \funpred^{\envreal}_{\typeone}$ and
$\sem{\envreal \imp \termone: \typeone} = \sem{\envreal \imp \termtwo: \typeone}$, 
then $\termone \in \funpred^{\envreal}_{\typeone}$.
The proof follows the same structure of Lemma~\ref{lemma:fundamental-lemma}, and thus 
we omit details here. 
\end{proof}

Using Lemma \ref{lemma:fundamental-lemma-logical-predicate} 
we immediately obtain the wished result.

\begin{mytheorem}[\ctheorem]
Let $\funsymbols$ be a collection of real-valued functions including 
projections and closed under function composition. Then, any 
$\STLambdarealfun$ term 
$\varone_1: \typereal, \hh, \varone_n: \typereal_n \imp \termone: \typereal$ 
denotes a function (from $\mathbb{R}^n$ to $\mathbb{R}$) in 
$\funsymbols$. That is, $\sem{\varone_1: \typereal, \hh, \varone_n: \typereal_n \imp \termone: \typereal} \in \funsymbols$.
\end{mytheorem}

\section{Automatic Differentiation}
\label{sect:automatic-differentiation}

In this section, we show how we can use open logical relations to 
prove the correctness of the automatic differentiation
algorithm of \cite{Efficient/AD/jones} (suitably adapted to our calculus). 

\emph{Automatic differentiation} 
\cite{BARTHOLOMEWBIGGS2000171,DBLP:journals/jmlr/BaydinPRS17,Griewank:2008:EDP:1455489} 
(AD, for short) is a family of techniques to 
efficiently
compute the \emph{numerical} (as opposed to \emph{symbolical}) 
derivative of a computer program denoting a real-valued function. 
Roughly speaking, 
AD acts on the code of a program by letting variables incorporate values for 
their derivative, and operators propagate derivatives according to the 
\emph{chain rule} of differential calculus \cite{spivak1971calculus}.
Due to its vast applications in machine learning (backpropagation~\cite{Rumelhart:1988:LRB:65669.104451} being an example of an AD technique) 
and, most notably, in deep learning~\cite{DBLP:journals/jmlr/BaydinPRS17}, 
AD is rapidly becoming a topic of interest in 
the programming language theory community, as witnessed by the new line of research 
called \emph{differentiable programming} 
(see, e.g., \cite{DBLP:journals/pacmpl/Elliott18,Efficient/AD/jones,DBLP:journals/pacmpl/BrunelMP20,DBLP:journals/pacmpl/AbadiP20} for some recent results on AD and programming language theory 
developed in the latter field).

AD comes in two modes, \emph{forward mode} (also called \emph{tangent mode}), 
and \emph{backward mode} (also called \emph{reverse mode}). 
These can be seen as different ways to compute the chain rule, the former by traversing 
the chain rule from inside to outside, while the latter from outside to inside. 

Here we are concerned with forward mode AD. More specifically, we consider the forward mode AD algorithm recently proposed 
in \cite{Efficient/AD/jones}. The latter is based on a
source-to-source program transformation extracting out of a program 
$\termone$ a new program $\de{\termone}$ whose evaluation simultaneously gives 
the result of computing $\termone$ and its derivative. This is achieved  
by augmenting the code of $\termone$ in such a way to handle 
\emph{dual numbers}\footnote{ 
Recall that a dual number~\cite{10.1112/plms/s1-4.1.381} 
is a pair of the form $(x,x')$, with 
$x,x' \in \mathbb{R}$. The first component, namely $x$, is subject to the usual real-number arithmetic, 
whereas the second component, namely $x'$, obeys to first-order differentiation arithmetic. 
Dual numbers are usually presented, in analogy with complex numbers, as 
formal sums of the form $x + x' \varepsilon$, where $\varepsilon$ is an abstract number 
(an infinitesimal) subject to the law $\varepsilon^2 = 0$.}.

The transformation roughly goes as 
follows:
expressions $\termtwo$ of type $\typereal$ are transformed into dual numbers, i.e. 
expressions 
$\termtwo'$
of type $\typereal \times \typereal$, where the first component of $\termtwo'$ gives the 
original value of $\termtwo$, and the second component of $\termtwo'$ gives 
the derivative of $\termtwo$.  
Real-valued function symbols are then extended to handle dual numbers by applying the chain rule,
while other constructors of the language are extended pointwise.

The algorithm of \cite{Efficient/AD/jones} has been studied by means 
of benchmarks but, 
to the best of the authors knowledge no formal proof of its correctness has 
been given so far. Differentiability being a first-order concept, open logical 
relations are thus a perfect candidate for such a job.

\paragraph{An AD Program Transformation} 
In the rest of this section, given a differentiable function 
$\funsymbol: \mathbb{R}^n \to \mathbb{R}$, we denote by 
$\partial_x \funsymbol: \mathbb{R}^n \to \mathbb{R}$ 
its partial derivative with respect to the variable $x$. 
Let $\derivfun$ be the collection of (real-valued) differentiable 
functions (in any variable), and let us fix a collection 
$\funsymbols$ of real-valued functions such that, for any 
$\funsymbol \in \derivfun$, both $f$ and $\partial_x \funsymbol$ belong to 
$\funsymbols$. Notice that since $\partial_x \funsymbol$ is not necessarily 
differentiable, in general $\partial_x \funsymbol \not \in \derivfun$. 

We begin recalling how the program transformation of \cite{Efficient/AD/jones} 
works on $\STLambdarealderiv$, the extension of $\STLambdareal$ with 
operators for functions in $\derivfun$. In order to define the derivative of a 
$\STLambdarealderiv$ expression, 
we first define an intermediate program transformation 
$\de{}: \STLambdarealderiv \to \STLambdarealfun$ such that:
$$
\envone \imp \termone: \typeone \implies 
\de{\envone} \imp \de{\termone}: \de{\typeone}.
$$
The action of $\de{}$ on types, environments, and expressions is defined 
in \autoref{figure:defintion-de}.
Notice that $\termone$ is an expression in $\STLambdarealderiv$, whereas 
$\de{\termone}$ is an expression in $\STLambdarealfun$. 

\begin{center}
\begin{figure}[htbp]
\fbox{
\begin{minipage}{.97\textwidth}
{\center
\begin{align*}
\de{\typereal} 
&= \typereal \times \typereal
&\de{(\cdot)} &= \cdot
\\
\de{(\typeone_1 \times \typeone_2)}
&= \de{\typeone_1} \times \de{\typeone_2}
& \de{(\varone: \typeone, \envone)} 
&= \devar{\varone}: \de{\typeone}, \de{\envone}
\\
\de{(\typeone_1 \to \typeone_2)}
&= \de{\typeone_1} \to \de{\typeone_2}
& &
\end{align*}
\vspace{-0.7cm}
\begin{align*}
\de{\makereal{r}} &= (\makereal{r}, \makereal{0})
&
\de{(\makereal{\funsymbol}(\termone_1, \hh, \termone_n))} 
&=
(\makereal{\funsymbol}(\fst{\de{\termone_1}}, \hh, \fst{\de{\termone_n}}),
\sum_{i = 1}^{n} \makereal{\partial_{x_i} \funsymbol}
(\fst{\de{\termone_1}}, \hh, \fst{\de{\termone_n}}) \mathbin{*} 
\snd{\de{\termone_i}})
\end{align*}
$\vspace{-0.6cm}$
\begin{align*}
\de{\varone} &= \devar{\varone} 
&
\de{(\abs{\varone}{\termone})} &= \abs{\devar{\varone}}{\de{\termone}}
&
\de{(\termtwo \termone)} &= (\de{\termtwo}) (\de{\termone})
&
\de{(\ith{\termone})} &= \ith{\de{\termone}}
&
\de{(\termone_1, \termone_2)} &= (\de{\termone_1}, \de{\termone_2})
\end{align*}
}
\end{minipage}}
\caption{Intermediate transformation $\de{}$}
\label{figure:defintion-de}
\end{figure}
\end{center}
Let us comment the definition of $\de{}$, beginning with its action on types.
Following the rationale behind forward-mode AD, the map $\de{}$ associates to the 
type $\typereal$ the product type 
$\typereal \times \typereal$, the first and second components of its inhabitants
being the original expression and its derivative, respectively.
The action on $\de{}$ on non-basic types is straightforward and it is designed 
so that the automatic differentiation machinery can handle higher-order expressions 
in such a way to guarantee correctness at real-valued function types. 

The action of $\de{}$ on the usual constructors of the $\lambda$-calculus is pointwise, 
although it is worth noticing that $\de{}$ associate to any variable 
$\varone$ of type $\typeone$ 
a new variable, which we denote by $\devar{\varone}$, of type $\de{\typeone}$. 
As we are going to see, if $\typeone = \typereal$, then $\devar{\varone}$ acts 
as a placeholder for a dual number.

More interesting is the action of $\de{}$ on real-valued constructors. 
To any numeral $\makereal{r}$, $\de{}$ associates the pair $\de{\makereal{r}} = 
(\makereal{r}, \makereal{0})$, the derivative of a number being zero. Let us now 
inspect the action of $\de{}$ on 
an operator $\makereal{\funsymbol}$ 
associated to $\funsymbol: \mathbb{R}^n \to \mathbb{R}$ (we treat $\funsymbol$ as 
a function in the variables $x_1, \hh, x_n$). 
The interesting 
part is the second component of $\de{(\makereal{\funsymbol}(\termone_1, \hh, 
\termone_n))}$, namely
$$
\sum_{i = 1}^{n} \makereal{\partial_{x_i} \funsymbol}
(\fst{\de{\termone_1}}, \hh, \fst{\de{\termone_n}}) \mathbin{*} 
\snd{\de{\termone_i}}
$$
where $\sum_{i=1}^{n}$ and $*$ denote the operators (of $\STLambdarealfun$) 
associated to summation and (binary) multiplication (for readability we omit 
the underline notation), and $\makereal{\partial_{x_i} \funsymbol}$ 
is the operator (of $\STLambdarealfun$) associated to partial derivative 
$\partial_{x_i} \funsymbol$ of $\funsymbol$ in the variable $x_i$. 
It is not hard to recognize that 
the above expression is nothing but an instance of the
\emph{chain rule}. 


Finally, we notice that if 
$\envone \imp \termone: \typeone$ is a (derivable) judgment in 
$\STLambdarealderiv$, then indeed $\de{\envone} \imp \de{\termone}: \de{\typeone}$ 
is a (derivable) judgment in $\STLambdarealfun$.

\begin{example}
Let us consider the binary function $\funsymbol(x_1,x_2) = \sin(x_1) + \cos(x_2)$. 
For readability, we overload the notation writing $\funsymbol$ in place of 
$\makereal{\funsymbol}$ (and similarly for $\partial_{x_i} \funsymbol$). 
Given expressions $\termone_1, \termone_2$, we compute 
$\de{(\sin(\termone_1) + \cos(\termone_2))}$. 
Recall that $\partial_{x_1} \funsymbol(x_1,x_2) = \cos(x_1)$ and 
$\partial_{x_2} \funsymbol(x_1,x_2) = - \sin(x_2)$. We have:
\begin{align*}
&\de{(\sin(\termone_1) + \cos(\termone_2))}
\\
&= (\sin(\fst{\de{\termone_1}}) + \cos(\fst{\de{\termone_2}}), 
  \partial_{x_1} \funsymbol(\fst{\de{\termone_1}}, \fst{\de{\termone_2}}) * 
  \snd{\de{\termone_1}} +
  \partial_{x_2} \funsymbol(\fst{\de{\termone_1}}, \fst{\de{\termone_2}}) * 
  \snd{\de{\termone_2}})
  \\
&= (\sin(\fst{\de{\termone_1}}) + \cos(\fst{\de{\termone_2}}), 
  \cos(\fst{\de{\termone_1}}) * \snd{\de{\termone_1}} 
  - \sin(\fst{\de{\termone_2}}) * \snd{\de{\termone_2}}).
\end{align*}
As a consequence, we see that
$
\de{(\lambda x.\lambda y. \sin(x) + \cos(y))} 
$ is
$$
\lambda \devar{x}.\lambda{\devar{y}}.
(\sin(\fst{\devar{x}}) + \cos(\fst{\devar{y}}), 
  \cos(\fst{\devar{x}}) * \snd{\devar{x}} 
  - \sin(\fst{\devar{y}}) * \snd{\devar{y}}).
$$
\end{example}

We now aim to define the derivative of an expression 
$\varone_1 : \typereal, \hh, \varone_n:\typereal \imp \termone: \typereal$ 
with respect to a variable $\varone$ (of type $\typereal$). 
In order to do so we first associate to any variable 
$\vartwo: \typereal$ its dual expression
$\dual{\varone}{\vartwo}: \typereal \times \typereal$ defined as:
\begin{align*}
\dual{\varone}{\vartwo}
&= \begin{cases}
    (\vartwo, \makereal{1}) &\text{ if } \varone = \vartwo 
    \\
    (\vartwo, \makereal{0}) &\text{ otherwise.}
    \end{cases}
\end{align*}
Next, we define for 
$\varone_1 : \typereal, \hh, \varone_n:\typereal \imp \termone: \typereal$
the derivative $\deriv{\varone}{\termone}$ of $\termone$ with respect to 
$\varone$ as:
$$
\deriv{\varone}{\termone} 
= \snd{\de{\termone}[\dual{x}{x_1}/\devar{x_1}, \hh, \dual{x}{x_n}/\devar{x_n} ]}
$$
Let us clarify this passage with a simple example.

\begin{example}
Let us compute the derivative of 
$x: \typereal, y: \typereal \imp \termone: \typereal$, where 
$\termone = x * y$. We first of all compute $\de{\termone}$, obtaining:
$$
\devar{x}: \typereal \times \typereal, \devar{y}: \typereal \times \typereal 
\imp ((\fst{\devar{x}}) * (\fst{\devar{y}}), (\fst{\devar{x}}) * (\snd{\devar{y}}) + 
(\snd{\devar{x}}) * (\fst{\devar{y}})): \typereal \times \typereal.$$
Observing that $\dual{x}{x} = (x, \makereal{1})$ and 
$\dual{x}{y} = (y, \makereal{0})$, 
we indeed obtain the wished derivative as:
$$
x: \typereal, y:\typereal \imp 
\snd{\de{\termone}
[\dual{x}{x}/\devar{x}, \dual{x}{y}/\devar{y}]}: \typereal
$$
For we have:
\begin{align*}
& \sem{x: \typereal, y:\typereal \imp 
\snd{\de{\termone}[\dual{x}{x}/\devar{x}, \dual{x}{y}/\devar{y}]}: \typereal}
\\
&= \sem{x: \typereal, y:\typereal \imp 
\snd{((\fst{(x, \makereal{1})}) * (\fst{(y, \makereal{0})}), (\fst{(x, \makereal{1})}) * 
(\snd{(y, \makereal{0})}) + 
(\snd{(x, \makereal{1})}) * (\fst{(y, \makereal{0})}))}: \typereal}
\\
&= \sem{x: \typereal, y:\typereal \imp 
\snd{(x * y, x * 0 + 1 * y)}: \typereal}
\\
&= \sem{x: \typereal, y:\typereal \imp y: \typereal}
\\
&= \partial_x \sem{x: \typereal, y: \typereal \imp x * y: \typereal}
\end{align*}
\end{example}

\begin{remark}
Notice that for $\envreal = \varone_1: \typereal_1, \hh, 
\varone_n: \typereal_n$ we 
have $\envreal \imp \dual{y}{x_i}: \de{\typereal}$ and 
$\envreal \imp \de{\termtwo}[\dual{y}{x_1}/\devar{\varone}_1, \hh, 
\dual{y}{x_n}/\devar{\varone}_n]: \de{\typeone}$, for any 
variable $y$ and $\envreal \imp \termtwo: \typeone$.
\end{remark}
\longv{
\begin{lemma}
  The following hold:
  \begin{varenumerate}
    \item $\forall y.\ \envreal \imp \dual{y}{x_i}: \de{\typereal}$.
    \item $\forall y.\ \envreal \imp \termtwo: \typeone \implies 
      \envreal \imp \de{\termtwo}[\devar{\varone}_1 := \dual{y}{x_1}, \hh, 
      \devar{\varone}_n := \dual{y}{x_n}]: \de{\typeone}$.
  \end{varenumerate}
\end{lemma}

  \begin{proof}
  Straightforward. For point $1$ simply observe that 
  $FV(\dual{y}{x_i}) = \{x_i\}$, whereas point $2$ follows since 
  $\envreal \imp \termtwo: \typeone$ implies 
  $\de{\envreal} \imp \de{\termtwo}: \de{\typeone}$, and 
  $\de{\envreal} = \devar{x}_1: \de{\typereal}, \hh, \devar{x}_1: \de{\typereal}$.
  \end{proof}
}


\paragraph{Open Logical relations for AD}

We have claimed the operation $\mathtt{deriv}$ to perform automatic differentiation 
of $\STLambdarealderiv$ expressions. By that we mean that once applied to 
expressions of the form
$\varone_1: \typereal, \hh, \varone_n: \typereal \imp \termone: \typereal$, 
the operation $\mathtt{deriv}$ can be used to compute 
the derivative 
of 
$\sem{\varone_1: \typereal, \hh, \varone_n: \typereal \imp \termone: \typereal}$. 
We now show how we can prove such a statement using open logical relations, 
this way providing a proof correctness of our AD program transformation.

We begin defining a logical relations $\relone$ between $\STLambdarealderiv$ 
and $\STLambdarealfun$ expressions. We design $\relone$ in such a way that 
(i) $\termone \relone \de{\termone}$ and (ii) if $\termone \relone \termtwo$, 
then indeed $\termtwo$ corresponds to the derivative of $\termone$. 
While (ii) essentially holds by definition, (i) requires some efforts 
in order to be proved. 

\begin{definition}[Open Logical Relation] 
\label{def:open-logical-relation-AD}
Let $\envreal = \varone_1: \typereal_1, \hh, 
\varone_n: \typereal_n$ be a fixed, arbitrary environment.
Define the family of relations 
$(\relone^{\envreal}_{\typeone})_{\envreal, \typeone}$ 
between $\STLambdarealderiv$ 
and $\STLambdarealfun$ expressions   
by induction 
on $\typeone$ as follows:
\begin{align*}
  \termone \relone^{\envreal}_{\typereal} \termtwo 
  &\iff 
  \begin{cases}
    \envreal \imp \termone: \typereal 
    \wedge
    \de{\envreal} \imp \termtwo: \typereal \times \typereal &
    \\
    \forall y: \typereal.\ 
    \sem{\envreal \imp 
    \fst{\termtwo[\dual{y}{x_1}/\devar{x_1}, \hh, \dual{y}{x_n}/\devar{x_n}]}
    : \typereal} = \sem{\envreal \imp \termone: \typereal} &
    \\
    \forall y:\typereal.\
    \sem{\envreal \imp  
    \snd{\termtwo[\dual{y}{x_1}/\devar{x_1}, \hh, \dual{y}{x_n}/\devar{x_n}]}: 
    \typereal} 
    = \partial_y \sem{\envreal \imp \termone: \typereal} & 
    \end{cases}
  \\
  \termone \relone^{\envreal}_{\typeone_1 \to \typeone_2} \termtwo 
  &\iff
   \begin{cases}
    \envreal \imp \termone: \typeone_1 \to \typeone_2 
    \wedge
    \de{\envreal} \imp \termtwo: \de{\typeone_1} \to \de{\typeone_2} &
    \\
    \forall \termthree, \termfour.\ 
    \termthree \relone^{\envreal}_{\typeone_1} \termfour \implies 
    \termone \termthree \relone^{\envreal}_{\typeone_2} \termtwo \termfour &
    \end{cases}
  \\
  \termone \relone^{\envreal}_{\typeone_1 \times \typeone_2} \termtwo 
  &\iff
   \begin{cases}
    \envreal \imp \termone: \typeone_1 \times \typeone_2 
    \wedge
    \de{\envreal} \imp \termtwo: \de{\typeone_1} \times \de{\typeone_2} &
    \\ 
    \forall i \in \{1,2\}.\ \ith{\termone} \relone^{\envreal}_{\typeone_i} \ith{\termtwo} &
    \end{cases}
\end{align*}
We extend 
$\relone^{\envreal}_{\typeone}$ 
to 
the family $(\relone^{\envone,\envreal}_{\typeone})_{\envone, \envreal, \typeone}$, 
where $\envone$ ranges over arbitrary environments (possibly containing 
variables of type $\typereal$), as follows: 
\begin{align*}
\termone \relone^{\envone, \envreal}_{\typeone} \termtwo
&\iff 
(\envone, \envreal \imp \termone: \typeone) \wedge
(\de{\envone}, \de{\envreal} \imp \termtwo: \de{\typeone}) \wedge
(\forall \substone, \substone'.\ 
\substone \relone^{\envone}_{\envreal} \substtwo \implies 
\termone \substone \relone^{\envreal}_{\typeone} \termtwo \substtwo) 
\end{align*}
where $\substone$, $\substtwo$ range over substitutions, and:
\begin{align*}
\substone \relone^{\envone}_{\envreal} \substtwo 
&\iff 
(\support{\substone} = \envone) \wedge 
(\support{\substtwo} = \de{\envone}) \wedge
(\forall (\varone: \typeone) \in \envone.\ 
\substone(\varone) \relone^{\envreal}_{\typeone} \substtwo(\devar{\varone})). 
\end{align*}
\end{definition}

Obviously, \autoref{def:open-logical-relation-AD} satisfies 
condition (ii) above. What remains to be done is to show that it 
satisfies condition (i) as well. 
In order to prove such a result, we first need to show that 
the logical relation respects the denotational semantics of 
$\STLambdarealderiv$.

\begin{lemma}
\label{lemma:expansion-open-log-rel}
Let $\envreal = x_1: \typereal, \hh, x_n: \typereal$. Then, the following hold:
\begin{align*}
  \termone' \relone^{\envreal}_{\typeone} \termtwo \wedge 
  \sem{\envreal \imp \termone: \typeone} = \sem{\envreal \imp \termone': \typeone} 
  &\implies \termone \relone^{\envreal}_{\typeone} \termtwo
  \\
  \termone \relone^{\envreal}_{\typeone} \termtwo' \wedge 
  \sem{\de{\envreal} \imp \termtwo': \de{\typeone}} = 
  \sem{\de{\envreal} \imp \termtwo: \de{\typeone}}
  &\implies \termone \relone^{\envreal}_{\typeone} \termtwo.
\end{align*}
\end{lemma}

\begin{proof}
A standard induction on $\typeone$. 
\longv{  We show how to prove the second one. 
  \begin{description}
    \item[Case 1.] Suppose $\typeone = \typereal$. In order to prove 
    $\termone \relone^{\envreal}_{\typeone} \termtwo$ we have to show:
    \begin{align*}
      \forall y: \typereal.\ 
      \sem{\envreal \imp 
      \fst{\termtwo[\devar{x_1} := \dual{y}{x_1}, \hh, \devar{x_n} := \dual{y}{x_n}]}
      : \typereal} &= \sem{\envreal \imp \termone: \typereal}
      \\
      \forall y:\typereal.\
      \sem{\envreal \imp  
      \snd{\termtwo[\devar{x_1} := \dual{y}{x_1}, \hh, \devar{x_n} := \dual{y}{x_n}]}: 
      \typereal} 
      &= \partial_y \sem{\envreal \imp \termone: \typereal}.
    \end{align*}
    We have:
    \begin{align*}
      & \sem{\envreal \imp  
      \ith{\termtwo[\devar{x_1} := \dual{y}{x_1}, \hh, \devar{x_n} := \dual{y}{x_n}]}: 
      \typereal} 
      \\
      & = \pi_i \comp
      \sem{\de{\envreal} \imp \termtwo: \de{\typereal}} \comp \prod_i \sem{x_i: \typereal 
      \imp \dual{y}{x_i}: \de{\typereal}}
      \\
      &= \pi_i \comp
      \sem{\de{\envreal} \imp \termtwo': \de{\typereal}} \comp \prod_i \sem{x_i: \typereal 
      \imp \dual{y}{x_i}: \de{\typereal}}
      \\
      &= \sem{\envreal \imp  
      \ith{\termtwo[\devar{x_1} := \dual{y}{x_1}, \hh, \devar{x_n} := \dual{y}{x_n}]}: 
      \typereal},
    \end{align*}
    so that the thesis follows from $\termone \relone^{\envreal}_{\typereal} \termtwo'$. 
    \item[Case 2.] Suppose $\typeone = \typeone_1 \to \typeone_2$. By very definition of 
      open logical relation, we have to show 
      $\termone \termfour \relone^{\envreal}_{\typeone_2} \termtwo \termfour'$, for all 
      $\termfour, \termfour'$ such that $\termfour \relone^{\envreal}_{\typeone_1} \termfour'$.
      Let $\termfour, \termfour'$ be such expressions. From 
      $\termone \relone^{\envreal}_{\typeone_1 \to \typeone_2} \termtwo'$ we infer 
      $\termone \termfour \relone^{\envreal}_{\typeone_2} \termtwo' \termfour'$. 
      We conclude the wished thesis by induction hypothesis, according to the 
      following identities:
      \begin{align*}
      \sem{\de{\envreal} \imp \termtwo \termfour' : \de{\typeone_2}}
      &= \mathsf{eval} \comp 
      \lan \sem{\de{\envreal} \imp \termtwo: \de{\typeone_1} \to \de{\typeone_2}},
      \sem{\de{\envreal} \imp \termfour': \de{\typeone_1}}
      \ran
      \\
      &=  \mathsf{eval} \comp 
      \lan \sem{\de{\envreal} \imp \termtwo': \de{\typeone_1} \to \de{\typeone_2}},
      \sem{\de{\envreal} \imp \termfour': \de{\typeone_1}}
      \ran
      \\
      &= \sem{\de{\envreal} \imp \termtwo' \termfour' : \de{\typeone_2}}.
      \end{align*}
     \item[Case 3.] If $\typeone = \typeone_1 \times \typeone_2$, then we proceed 
      following the same patter of case $2$.
  \end{description}
}
\end{proof}

We are now ready to state and prove the main result of this section.

\begin{lemma}[Fundamental Lemma]
\label{lemma:fundamental-lemma}
For all environments $\envone, \envreal$ and 
for any expression $\envone, \envreal \imp \termone: \typeone$, 
we have $\termone \relone^{\envone, \envreal}_{\typeone} \de{\termone}$. 
\end{lemma}

\begin{proof}
We prove the following statement, by induction on $\termone$: 
\begin{align*}
\forall \termone.\ \forall \typeone.\ \forall \envone, \envreal.\  
(\envone, \envreal \imp \termone: \typeone \implies 
\termone \relone^{\envone, \envreal}_{\typeone} \de{\termone}).
\end{align*}
We show only the most relevant cases.

\begin{varitemize}
  \item Suppose $\termone$ is a variable $x$. We distinguish 
    whether $x$ belongs to $\envone$ or $\envreal$.
    \begin{varenumerate}
      \item Suppose $(x: \typereal) \in \envreal$. We have to 
        show $\varone \relone^{\envone, \envreal}_{\typereal} \devar{\varone}$, i.e. 
        \begin{align*}
          \sem{\envreal \imp \fst{\devar{\varone}[\dual{y}{x}/\devar{\varone}]}: \typereal}
          &= \sem{\envreal \imp \varone: \typereal}
          \\
          \sem{\envreal \imp \snd{\devar{\varone}[\dual{y}{x}/\devar{\varone}]}: \typereal}
          &= \partial_{y} \sem{\envreal \imp \varone: \typereal}
        \end{align*} 
        for any variable $y$ (of type $\typereal$). 
        The first identity obviously holds as 
        \begin{align*}
        \sem{\envreal \imp \fst{\devar{\varone}[\dual{y}{x}/\devar{\varone}]}: 
        \typereal}  
        = \sem{\envreal \imp \fst{\devar{\varone}[(\varone, b)/\devar{\varone}]}: 
        \typereal}
        = \sem{\envreal \imp x: \typereal},
        \end{align*}
        where $b \in \{\makereal{0}, \makereal{1}\}$.
        For the second identity we distinguish whether $y = \varone$ or 
        $y \neq \varone$. In the former case we have $\dual{y}{x} = (x, \makereal{1})$, 
        and thus:
        \begin{align*}
        \sem{\envreal \imp \snd{\devar{\varone}[\dual{y}{x}/\devar{\varone}]}: 
        \typereal}  
        = \sem{\envreal \imp \makereal{1}: \typereal}
        = \partial_y \sem{\envreal \imp y: \typereal}.
        \end{align*}
        In the latter case we have $\dual{y}{x} = (x, \makereal{0})$, 
        and thus:
        \begin{align*}
        \sem{\envreal \imp \snd{\devar{\varone}[\dual{y}{x}/\devar{\varone}]}: 
        \typereal}  
        = \sem{\envreal \imp \makereal{0}: \typereal}
        = \partial_y \sem{\envreal \imp x: \typereal}.
        \end{align*}
      \item Suppose $(x: \typeone) \in \envone$. We have to 
        show $\varone \relone^{\envone, \envreal} \devar{\varone}$, i.e. 
        $\substone(\varone) \relone^{\envreal}_{\typeone} \substtwo(\devar{\varone})$, 
        for all substitutions $\substone, \substtwo$ such that 
        $\substone \relone^{\envone}_{\envreal} \substtwo$. Since $x$ belongs to 
        $\envone$, we are trivially done.
    \end{varenumerate}
  \item Suppose $\termone$ is $\abs{\varone}{\termtwo}$, so that 
    we have
    \[
    \infer{\envone, \envreal \imp \abs{\varone}{\termtwo}:\typeone_1 \to \typeone_2}
    {
    \deduce[\vdots]{\envone, \envreal, \varone: \typeone_1 \imp \termtwo: \typeone_2}{}
    }
    \]
    for some types $\typeone_1, \typeone_2$. 
    As $\varone$ is bound in $\abs{\varone}{\termtwo}$, without loss of 
    generality we can assume $(\varone: \typeone_1) \not \in \envone \cup \envreal$.
    Let $\Delta = \envone, \varone: \typeone_1$, so that we have 
    $\Delta, \envreal \imp \termtwo: \typeone_2$, and thus 
    $\termtwo \relone^{\Delta, \envreal}_{\typeone_2} \de{\termtwo}$, 
    by induction hypothesis. By very definition of open logical relation, 
    we have to prove that for arbitrary $\substone, \substtwo$ such that 
    $\substone \relone^{\envone}_{\envreal} \substtwo$, we have 
    $$
    \abs{\varone}{\termtwo \substone} \relone^{\envreal}_{\typeone_1 \to \typeone_2} 
    \abs{\devar{\varone}}{(\de{\termtwo})\substtwo},
    $$ 
    i.e. 
    $
    (\abs{\varone}{\termtwo \substone})\termthree \relone^{\envreal}_{\typeone_2} 
    (\abs{\devar{\varone}}{(\de{\termtwo})\substtwo})\termfour
    $,
    for all $\termthree \relone^{\envreal}_{\typeone_1} \termfour$. Let us fix a pair 
    $(\termthree,\termfour)$ as above. By Lemma \ref{lemma:expansion-open-log-rel}, 
    it is sufficient to show 
    $$
    \subst{(\termtwo \substone)}{\varone}{\termthree} 
    \relone^{\envreal}_{\typeone_2} 
    \subst{((\de{\termtwo})\substtwo)}{\devar{\varone}}{\termfour}.
    $$
    Let $\substone', \substtwo'$ be the substitutions defined as follows:
    \begin{align*}
    \substone'(y) 
    &= 
    \begin{cases}
    \termthree           & \text{ if }  y = \varone 
    \\
    \substone(y) &\text{ otherwise }
    \end{cases}
    &
    \substtwo'(y)
    &= 
    \begin{cases}
    \termfour            & \text{ if } y = \devar{x}
    \\
    \substtwo(y)  &\text{ otherwise. }
    \end{cases}
    \end{align*} 
    \longv{
    \begin{description}
    \item[Claim.] $\substone' \relone^{\Delta}_{\envreal} \substtwo'$.
    \item[Proof.]
    Since $\substone \relone^{\envone}_{\envreal} \substtwo$, we have 
    $\support{\substone} = \envone$ and $\support{\substtwo} = \de{\envone}$. 
    Therefore
    \begin{align*}
    \support{\substone'} 
    &= \support{\substone} \cup \{(\varone: \typeone_1)\}
    &
    \support{\substtwo'} 
    &= \support{\substtwo} \cup \{(\devar{\varone}: \de{\typeone_1})\}
    \\
    &= \envone \cup \{(\varone: \typeone_1)\}
    &
    &= \de{\envone} \cup \{(\devar{\varone}: \de{\typeone_1})\}
    \\
    &= \Delta
    &
    &= \de{\Delta}.
    \end{align*}
    It remains to prove that for any $(y: \typeone) \in \Delta$ we have 
    $\substone'(y) \relone^{\envreal}_{\typeone} \substtwo'(\devar{y})$. 
    Since $\substone \relone^{\envone}_{\envreal} \substtwo$, this is true 
    for any $(y: \typeone) \in \envone$. Therefore, it remains to prove 
    the case for $(x: \typeone_1)$, i.e. $k \relone^{\envreal}_{\typeone_1} k'$, 
    which holds by hypothesis. 
    \end{description}
    From $\substone' \relone^{\Delta}_{\envreal} \substtwo'$ and
    $\termtwo \relone^{\Delta, \envreal}_{\typeone_2} \de{\termtwo}$ 
    (recall that the latter follows by induction hypothesis) we infer 
    $$
    \termtwo \substone' \relone^{\envreal}_{\typeone_2} (\de{\termtwo})\substtwo'
    $$
    by 
    very definition of open logical relation.
    }
    \shortv{
    It is easy to see that $\substone' \relone^{\Delta}_{\envreal} \substtwo'$, 
    so that by 
    $\termtwo \relone^{\Delta, \envreal}_{\typeone_2} \de{\termtwo}$ 
    (recall that the latter follows by induction hypothesis) we infer 
    $
    \termtwo \substone' \relone^{\envreal}_{\typeone_2} (\de{\termtwo})\substtwo',
    $
    by 
    very definition of open logical relation.
    }
    As a consequence, the thesis is proved if we show 
    \begin{align*}
    \subst{(\termtwo \substone)}{\varone}{\termthree}
    &= \termtwo \substone'
    & 
    \subst{((\de{\termtwo})\substtwo)}{\devar{\varone}}{\termfour}
    &= (\de{\termtwo})\substtwo'.
    \end{align*}
    The above identities hold if $\varone \not \in FV(\substone(y))$ 
    and $\devar{\varone} \not \in FV(\substtwo(\devar{y}))$, for any $(y: \typeone) \in 
    \envone$. This is indeed the case, 
    since $\substone(y) \relone^{\envreal}_{\typeone} \substtwo(\devar{y})$ 
    implies $\envreal \imp \substone(y): \typeone$ and 
    $\de{\envreal} \imp \substtwo(\devar{y}): \de{\typeone}$, 
    and $\varone \not \in \envreal$ (and thus $\devar{\varone} \not \in 
    \de{\envreal}$).
\longv{
    \item Suppose $\termone$ to be $\termtwo \termfour$, so that 
    we have
    \[
    \infer{\envone, \envreal \imp \termtwo \termfour: \typeone_2}
    {
    \deduce[\vdots]{\envone, \envreal \imp \termtwo: \typeone_1 \to \typeone_2}{}
    &
    \deduce[\vdots]{\envone, \envreal \imp \termfour: \typeone_1}{}
    }
    \]
    for some types $\typeone_1, \typeone_2$. 
    By induction hypothesis we know 
    $\termtwo \relone^{\envone, \envreal}_{\typeone_1 \to \typeone_2} \de{\termtwo}$ 
    and $\termfour \relone^{\envone, \envreal}_{\typeone_1}$, from which the 
    thesis straightforwardly follows, since $\de{(\termtwo\termfour)} = 
    (\de{\termtwo})(\de{\termfour})$.
    \item Suppose $\termone$ to be $(\termone_1, \termone_2)$, so that 
    we have
    \[
    \infer{\envone, \envreal \imp (\termone_1, \termone_2):  \typeone_1 \times \typeone_2}
    {
    \deduce[\vdots]{\envone, \envreal \imp \termone_1: \typeone_1}{}
    &
    \deduce[\vdots]{\envone, \envreal \imp \termone_2: \typeone_2}{}
    }
    \]
    for some types $\typeone_1, \typeone_2$. We proceed following the same patter of Case 
    2, hence relying on \ref{lemma:expansion-open-log-rel}.
    \item If $\termone = \ith{\termtwo}$, then we proceed as in Case 3.
}
\end{varitemize}
\end{proof}

A direct application of Lemma~\ref{lemma:fundamental-lemma}
allows us to conclude the correctness of the program transformation 
$\mathtt{D}$. In fact, given a first-order term 
$\envreal \imp \termone: \typereal$, with 
$\envreal = \varone_1: \typereal, \hh, \varone_n: \typereal$, 
by Lemma~\ref{lemma:fundamental-lemma} we have 
$\termone \relone^{\envreal}_{\typereal} \de{\termone}$, and thus
$$
\partial_y \sem{\envreal \imp \termone: \typereal}
=
\sem{\envreal \imp  
    \snd{\de{\termone}[\dual{y}{x_1}/\devar{x_1}, \hh, \dual{y}{x_n}/\devar{x_n}]}: 
    \typereal},
$$ for any 
real-valued variable $y$, meaning that $\de{\termone}$ indeed 
computes the partial derivative of $\termone$.

\begin{theorem}
\label{theorem:olr-ad}
For any term $\envreal \imp \termone: \typereal$
as above, the term 
$\de{\envreal} \imp \de{\termone}: \de{\typereal}$ computes the 
partial derivative of $\termone$, in the sense that for any variable $\vartwo$ 
we have
$$
\partial_{\vartwo}\sem{\envreal \imp \termone: \typereal} = 
\sem{\envreal \imp  
\snd{\de{\termone}[\dual{y}{x_1}/\devar{x_1}, \hh, \dual{y}{x_n}/\devar{x_n}]}: \typereal}.
$$
\end{theorem}

\section{On Refinement Types and Local Continuity}\label{sect:continuity}
In Section~\ref{section:conservation-theorem}, we exploited open logical
relations to establish a containment theorem for the calculus $\STLambdarealfun$, 
i.e. the calculus $\STLambdareal$  
extended with real-valued functions belonging to a set $\funsymbols$ containing 
projections and closed under function composition. Since 
the collection $\contfun$ of (real-valued) \emph{continuous} functions 
satisfies both constraints, 
Theorem~\ref{thm:conservation-theorem} allows us to conclude that 
all first order terms of $\STLambdarealcont$ represent continuous functions. 

The aim of the present section is the development of a framework to prove 
continuity properties of programs in a calculus that goes beyond $\STLambdarealcont$. Accordingly, (i) 
we do not restrict our analysis to calculi having operators representing 
continuous real-valued functions only,
but consider operators for arbitrary real-valued functions, and (ii)
we add to our calculus an if-then-else 
construct whose
static semantics is captured by the following rule:
\[
\infer{\envone \imp \ifth{\termone}{\termtwo}{\termthree}: \typeone}
{\envone \imp \termone: \typereal 
&
\envone \imp \termtwo: \typeone 
&
\envone \imp \termthree: \typeone}
\]
The intended dynamic semantics of the term
$\ifth{\termone}{\termtwo}{\termthree}$ is the same as the one of
$\termtwo$ whenever $\termone$ evaluates to any real number $r \neq 0$ and the same as the
one of $\termthree$ if it evaluates to $0$.
\begin{center}
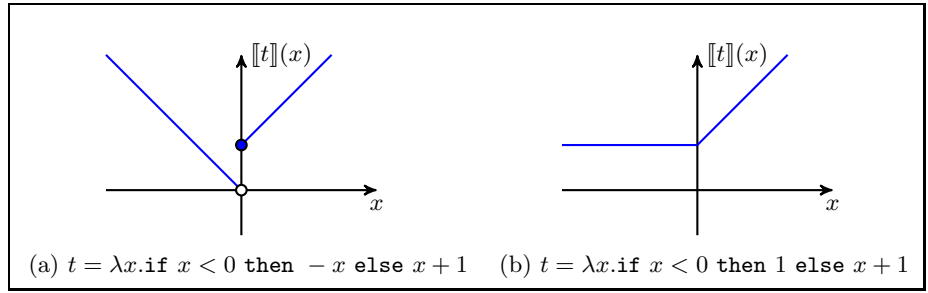
\begin{figure}[htpb]
\fbox{
\begin{minipage}{.97\textwidth}
\centering
\begin{subfigure}[b]{0.5\textwidth}
\centering
\vspace{0.3cm}
\begin{tikzpicture}[scale=0.6,
    thick,
    >=stealth',
    dot/.style = {
      draw,
      circle,
      inner sep = 0pt,
      minimum size = 4pt
    }
  ]
  \coordinate (O) at (0,0);
  \draw[->] (-3,0) -- (3,0) coordinate[label = {below:$x$}] (xmax);
  \draw[->] (0,-1) -- (0,3) coordinate[label = {right:$\sem \termone (x)$}] (ymax);
  \draw[color=blue] (0,1) -- (2,3);
  \draw[color=blue] (-3,3) -- (0,0);
  \node[dot, fill=white] at (0,0){};
  \node[dot, fill=blue] at (0,1){};
\end{tikzpicture}
 \caption{$\termone = \abs \varone {\ifth{x < 0}{-x}{x+1}}$}\label{fig:non_continuous_program}
    \end{subfigure}%
    ~ 
    \begin{subfigure}[b]{0.5\textwidth}
        \centering
       \begin{tikzpicture}[scale=0.6,
    thick,
    >=stealth']
  \coordinate (O) at (0,0);
  \draw[->] (-3,0) -- (3,0) coordinate[label = {below:$x$}] (xmax);
  \draw[->] (0,-1) -- (0,3) coordinate[label = {right:$\sem \termone (x)$}] (ymax);
  \draw[color=blue] (0,1) -- (2,3);
  \draw[color=blue] (-3,1) -- (0,1);
\end{tikzpicture}
\caption{$\termone = \abs \varone {\ifth{x < 0}{1}{x+1}}$}\label{fig:continuous_program}
\end{subfigure}
    \end{minipage}}
\caption{Simply typed first-order programs with branches}\label{fig:programs_with_branches}
  \end{figure}
\end{center}

Notice that the crux of the problem we aim to solve is given by the 
presence of the if-then-else construct. Indeed, independently of point (i),
such a construct breaks the global continuity of programs, as illustrated in
Figure~\ref{fig:non_continuous_program}. As a consequence we are
forced to look at \emph{local} continuity properties, instead: for
instance we can say that the program of
Figure~\ref{fig:non_continuous_program} is continuous both on
$\RR_{<0}$ and $\RR_{\geq 0}$. Observe that guaranteeing local continuity allows us (up to a certain point) to recover the ability of approximating the output of a program by approximating its input. Indeed, if a program $\termone : \typereal \times \ldots \times \typereal \to \typereal$ is \emph{locally continuous} on a subset $X$ of $\RR^n$, it means the value of $\termone \termtwo$ (for some input $\termtwo$) can be approximated by passing as argument to $\termone$ a family $(\termtwo_n)_{n \in \NN}$ of approximations of $\termtwo$, \emph{as long as} both $\termtwo$ and all the $(\termtwo_n)_{n \in \NN}$ are indeed elements of $X$. Notice that the continuity domains we are interested in are not necessary open sets: we could for instance be interested in functions that are continuous on the unit circle, i.e.~the points $\{(a,b) \mid a^2 + b^2 = 1\} \subseteq \RR^2$. For this reason we will work with the notion of \emph{sequential} continuity, instead of the usual topological notion of continuity. It must be observed, however, that these two notions coincide as soon as the continuity domain $X$ is actually an open set. 
\begin{definition}[Sequential Continuity]
Let $f : \RR^n \to \RR$, and $X$ be any subset of  $\RR^n$. We say that $f$ is \emph{(sequentially) continuous} on $X$ if for every $x \in X$, and for every sequence $(x_n)_{n \in \NN}$ of elements of $X$ such that $\lim_{n \to \infty} x_n = x$, it holds that $\lim_{n \to \infty} f(x_n) = f(x)$.
  \end{definition}

In~\cite{ChaudhuriGL10}, Chaudhuri et al.
introduced a logical system designed to guarantee local continuity
properties on programs in an \emph{imperative} (first-order) 
programming language
with conditional branches and loops. In this section, we develop
a similar system in the setting of a \emph{higher-order functional language}
with an if-then-else construct, and we use open logical relations to prove the soundness of our system.
This witnesses, on yet another situation, the versatility of open logical relations. Compared to~\cite{ChaudhuriGL10}, we somehow generalize from a result on programs built from only first-order constructs and primitive functions, to a containment result for programs built  using also higher-order constructs.

We however mention that, although our system is inspired by the work of 
Chaudhuri at al., there are significant differences between the two, 
even at the first-order level. The consequences these differences have on the expressive power of our systems are twofold:  
\begin{enumerate}[label=\roman*)]
  \item On one hand, while inferring continuity on some domain $X$ of a program of the form $\ifth \termone \termtwo \termthree$, we have more flexibility than~\cite{ChaudhuriGL10} for the domains of continuity of $\termtwo$ and $\termthree$. To be more concrete, let us consider the program:
    $\abs \varone(\ifth {(x>0)}{0}{(\ifth {x=4}{1}{0})}) $, which is continuous on $\RR$ even though the second branch is continuous on $\RR_{\leq 0}$, but not on $\RR$. We are able to show in our system that this program is indeed continuous on \emph{the whole} $\RR$, while Chaudhuri et al. cannot do the same in their system for the corresponding imperative program: they ask the domain of continuity of \emph{each} of the two branches to \emph{coincide} with the domain of continuity of the whole program.
    \item On the other hand, the system of Chaudhuri at al. allows one to express continuity along a restricted set of variables, which we cannot do. To illustrate this, let us look at the program:  $\abs {\varone, \vartwo} \ifth {(\varone = 0)}{(3 * \vartwo)}{(4 * \vartwo)} $: along the variable $\vartwo$, this program is continuous on the whole of $\RR$. Chaudhuri et al. are able to express and prove this statement in their system, while we can only say that for every real $a$, this program is continuous on the domain $\{a\} \times \RR$. 
\end{enumerate}

For the sake of simplicity, 
it is useful to slightly simplify our calculus; the ideas we present here, however, would still be valid in a more general setting, but that would make the presentation and proofs more involved. As usual, let 
$\funsymbols$ be a collection of real-valued functions. 
We consider the restriction of the calculus $\STLambdarealfun$ 
obtained by considering types of the form 
\begin{align*}
\typeone &\bnf \typereal \mid \rho 
& 
\rho &\bnf \rho_1 \times \cc \times \rho_n \times 
\underbrace{\typereal \times \cc \times \typereal}_{m\text{-times}} \to \typeone
\end{align*}
only. For the sake of readability, we employ the notation
$(\rho_1 \hh, \rho_n, \typereal, \hh, \typereal) \to \typeone$ 
in place of $\rho_1 \times \cc \times \rho_n \times 
\typereal \times \cc \times \typereal \to \typeone$. 
We also overload the notation and keep indicating the resulting calculus as 
$\STLambdarealfun$. 
Nonetheless, the reader should keep in mind that from now on, whenever referring 
to a $\STLambdarealfun$ term, we are tacitly referring to a term typable according 
to the restricted type system, but that can indeed contain conditionals.

Since we want to be able to talk about \emph{composition properties} of locally continuous programs, we actually need to talk not only about the points where a program is continuous, but also about the \emph{image} of this continuity domain. In higher-order languages, a well-established framework for the latter kind of specifications is   the one of \emph{refinement types}, that have been first introduced by~\cite{Freeman:1991:RTM:113445.113468} for \textsf{ML} types: the basic idea is to annotate an existing type system by logical formulas, with the aim of being more precise about the underlying program's behaviors than in simple types. Here, we are going to adapt this framework by replacing the image annotations provided by standard refinement types with \emph{continuity annotations}.

 \subsection{A Refinement Type System Ensuring Local Continuity.}\label{sec:cont}
 
Our refinement type system is developed on top of the simple type system of 
Section~\ref{sect:the_playground} (actually, on the simplification of such a system 
we are considering in this section). We first need to introduce a set of logical formulas which talk about $n$-uples of real numbers, and which we use as annotations in our refinement types. We consider a set $\variables$ of logical variables, and we construct formulas as follows:
  \begin{align*}
    & \psi, \phi \in \mathcal L \bnf \top \midd (e \leq e) \midd \psi \wedge \phi \midd \neg \psi, \\
   & e \in \exprs \bnf \alpha \midd a \midd f(e, \ldots, e) \qquad \text{ with } \alpha \in \variables, a \in \RR, f: \RR^n \to \RR
  \end{align*}
  
Recall that with the connectives in our logic, we are able  to encode logical disjunction and implication, and as customary, we write $\phi \Rightarrow \psi$ for 
$\neg \phi \vee \psi$.
A \emph{real assignment} is a partial map $\sigma: \variables \to \RR$. When $\sigma$ has finite support, we sometimes specify $\sigma$ by writing $(\varlone_1 \mapsto \sigma(\varlone_1), \ldots, \varlone_n \mapsto \sigma(\varlone_n))$.  We note $\sigma \models \phi$ when $\sigma$ is defined on the variables occuring in $\phi$, and moreover the real formula obtained when replacing along $\sigma$ the logical variables of $\phi$ is true. We write $\models \phi$ when $\sigma \models \phi$ always holds, independently on $\sigma$.

We can associate to every formula the subset of $\RR^n$ consisting of all points where this formula holds: more precisely, if $\phi$ is a formula, and 
$X = \alpha_1, \ldots, \alpha_n$ is a list of logical variables such that $\vars \phi \subseteq X$, we call \emph{truth domain of $\phi$ w.r.t. $X$} the set:
 $$\domain \phi^{X} =  \{(a_1, \ldots, a_n) \in \RR^n \mid  (\alpha_1 \mapsto a_1, \ldots, \alpha_n \mapsto a_n) \models \phi \}. $$
 
We are now ready to define the language of refinement types, which can be seen as simple types annotated by logical formulas.
  The type $\typereal$ is annotated by logical \emph{variables}: this way we obtain \emph{refinement real types} of the form $\{\alpha \in \typereal\}$. The crux of our refinement type system consists in the annotations we put \emph{on the arrows}.
We introduce two distinct refined
 arrow constructs, depending on the shape of the target type: more
 precisely we annotate the arrow of a type $(T_1, \ldots,
 T_n )\to \typereal$ with \emph{two} logical formulas, while we
 annotate $(T_1, \ldots, T_n )\to H$ (where $H$ is an higher-order type) with only \emph{one} logical
 formula. This way, we obtain refined arrow types of the form $(T_1, \ldots, T_n) \torr \psi \phi \{\alpha \in \typereal\}$, and  $(T_1, \ldots, T_n) \tor \psi H$: in both cases the formula $\psi$ specifies the continuity domain, while the formula $\phi$ is an \emph{image annotation} used only when the target type is ground. The intuition is as follows: a program of type $(H_1, \ldots, H_n, \{\alpha_1 \in \typereal\}, \ldots, \{\alpha_n \in \typereal\}) \torr \psi \phi \{\alpha \in \typereal\}$ uses its real arguments continuously on the domain specified by the formula $\psi$ (w.r.t $\alpha_1, \ldots, \alpha_n$), and this domain is sent into the domain specified by the formula $\phi$ (w.r.t. $\alpha$). Similarly,  a program of the type $(T_1, \ldots, T_n) \tor \psi H$ has its real arguments used in a continuous way on the domain specified by $\psi$, but it is not possible anymore to specify an image domain, because $H$ is higher-order.

 The general form of our refined types is thus as follows:
 \begin{align*}
  & \qquad T \bnf H \midd F; \qquad  F \bnf \{\alpha \in \typereal\}; 
  \\ & \qquad 
   H \bnf  (H_1, \ldots, H_m, F_1, \ldots,F_n) \tor \psi H \midd (H_1, \ldots, H_m, F_1, \ldots,F_n) \torr \psi \phi F 
  \end{align*}
  with $n+m >0$, $\vars \phi \subseteq \{ \alpha \}, \, \vars{\psi} \subseteq \{\alpha_1, \ldots, \alpha_n\}$  when $F = \{ \alpha \in \typereal\}$, 
 $F_i = \{ \alpha_i \in \typereal\}$, 
 and  the $(\alpha_i)_{1 \leq i \leq n}$ are distinct.
  We take refinement types up to renaming of logical variables. If $T$ is a refinement type, we write $\forget T$ for the simple type we obtain by forgetting about the annotations in $T$.
  \begin{example}
    We illustrate in this example the intended meaning of our refinement types.
    \begin{varitemize}
    \item We first look at how to refine $\typereal \to \typereal$:
      those are types of the form 
      $\{\alpha_1 \in \typereal\} \torr{\phi_1}{\phi_2} \{\alpha_2 \in \typereal\}$. 
      The intended inhabitants of
      these types are the programs
    $\termone: \typereal \to \typereal$ such that \begin{enumerate*}[label=\roman*), 
        itemjoin={{; }},itemjoin*={{; and }}] \item $\sem \termone$ is continuous on the truth domain of $\phi_1$ \item $\sem \termone$ sends the truth domain of $\phi_1$ into the truth domain of $\phi_2$.
      \end{enumerate*}
      As an example, $\phi_1$ could be $(\alpha_1 < 3)$, and $\phi_2$ could be $(\alpha_2 \geq 5)$. An example of a program having this type is $\termone = \abs \varone ( \makereal{5} + \makereal{f}(\varone))$, where $f: \RR \to \RR$ is defined as $f(a) = \begin{cases}\frac 1 {3-a} \text{ when }a < 3 \\ 0 \text{ otherwise}\end{cases},$ and moreover we assume that $\{f, + \}\subseteq \funsymbols$.
    \item We look now at the possible refinements of $\typereal \to
      (\typereal \to \typereal)$: those are of the form $\{\alpha_1
      \in \typereal\} \tor{\theta_1}{ (\{\alpha_2 \in
        \typereal \} \torr{\theta_2}{\theta_3} \{\alpha_3
        \in \typereal\})} $. The intended inhabitants of
      these types are the programs $\termone: \typereal \to (\typereal
      \to \typereal)$ whose interpretation function $(x,y)
      \in \RR^2 \mapsto \sem{\termone}(x)(y)$ sends continously
      $\domain {\theta_1}^{\alpha_1} \times \domain
          {\theta_2}^{\alpha_2}$ into $\domain {\theta_3}^{\alpha_3}$. As an example, consider $\theta_1 = (\alpha_1 < 1)$, $\theta_2 = (\alpha_2 \leq 3)$, and $\theta_3 = (\alpha_3 >0)$. An example of a program having this type is $\abs {\varone_1} {\abs {\varone_2} \makereal f(x_1 * x_2)}$ where we take $f$ as above.
\end{varitemize}
    \end{example}
  A refined typing context $\Gamma$ is a list $\varone_1:T_1, \ldots, \varone_n: T_n$, where each $T_i$ is a refinement type. In order to express continuity constraints, we need to \emph{annotate} typing judgments by logical formulas, in a similar way as what we do for arrow types. More precisely, we consider two kinds of refined typing judgments: one for terms of ground type, and one for terms of higher-order type:
  $$\Gamma \imprh \psi t:H \qquad \Gamma \imprr \psi \phi t:F. $$

  \subsection{Basic Typing Rules}
  We first consider the refined typing rules for the fragment of our
  language which excludes conditionals: they are given
  in Figure~\ref{fig:refined_typing_system}.  We  illustrate them by way of a series of examples.
\begin{center}
  \begin{figure}[htpb]
    \fbox{
      \begin{minipage}{0.97 \textwidth}
        \centering
    \begin{center}
        $$\AxiomC{\phantom{$ \models \theta \Rightarrow \theta'$}}
         \LeftLabel{var-H}
        \UnaryInfC{$ \Gamma, x: H {\imprh \psi} x: H $}
        \DisplayProof
 \qquad
 \AxiomC{$ \models \theta \Rightarrow \theta'$ }
 \LeftLabel{var-F}
\UnaryInfC{$ \Gamma, \varone: \{ \alpha \in \typereal\}  {\imprr \theta {\theta'}} {\varone: \{ \alpha \in \typereal\}}$}
\DisplayProof
$$
\vspace{0.2cm}
$$\AxiomC{$
  \begin{array}{l}
   f  \in \funsymbols \text{ is continuous on } \domain {\theta'_1 \wedge \ldots \wedge \theta'_n}^{\alpha_1 \ldots \alpha_n} \\
     f(\domain {\theta'_1 \wedge \ldots \wedge \theta'_n}^{\alpha_1 \ldots \alpha_n}) \subseteq \domain {\theta'}^\beta
\end{array}$
      }
\AxiomC{ $\Gamma {\imprr \theta {\theta'_i}} \termone_i : \{\alpha_i \in \typereal \}  $}
\LeftLabel{Rf}
\BinaryInfC{$ \Gamma {\imprr \theta {\theta'}} \makereal f (\termone_1 \ldots \termone_n ):  \{\beta \in \typereal\} $}
\DisplayProof
$$
\vspace{0.2cm}
$$
\AxiomC{$ \Gamma, \varone_1: T_1, \ldots, \varone_n:T_n {\imprh {\psi(\eta)}} \termone: T$}
\AxiomC{$\models \psi_1 \wedge \psi_2 \Rightarrow \psi$}
\LeftLabel{abs}
\BinaryInfC{$ \Gamma {\imprh {\psi_2}} \abs{(\varone_1, \ldots, \varone_n)} \termone : (T_1, \ldots, T_n) \tor{\psi_1(\eta)} T$}
\DisplayProof
$$
\vspace{0.2cm}
$$
\AxiomC{$\begin{array}{l}
  (\Gamma {\imprh \phi} \termtwo_i: H_i)_{1 \leq i \leq m}
    \\ \Gamma   {\imprh \phi} \termone:  (H_1, \ldots, H_m, F_1, \ldots,F_n) \tor {\theta(\eta)} T
  \end{array}$}
\AxiomC{$
  \begin{array}{l}
    \models \theta_1 \wedge \ldots \wedge \theta_n \Rightarrow \theta \\
   (\Gamma {\imprr \phi {\theta_j}} \termthree_j:F_j)_{1 \leq j \leq m}
   \end{array}
  $}
\LeftLabel{app}
\BinaryInfC{$ \Gamma {\imprh {\phi(\eta)}}  \termone (\termtwo_1, \ldots, \termtwo_m, \termthree_1, \ldots, \termthree_m) : T $}
\DisplayProof
$$
 \end{center}
   By abuse of notation $\psi(\eta)$ should be read as $\psi$ when $T$ is a higher-order type, and as $\psi \leadsto \eta$ when $T$ is a ground type.
 \end{minipage}}
 \caption{Typing Rules}\label{fig:refined_typing_system}
\end{figure}
\end{center}
 
  \begin{example}
    We first look at the typing  rule var-F: \RAPHAELLE{if $\theta$ implies $\theta'$, then the variable $x$\textemdash that, in semantics terms, do the projection of the context $\Gamma$ to one of its component\textemdash sends continuously the truth domain of $\Theta$ into the truth domain of $\Theta'$.}
    Using this rule
    we can, for instance, derive the following judgment:
    \begin{equation}
        \varone: \{ \alpha \in \typereal  \}, \vartwo: \{\beta \in \typereal  \}  {\imprr {(\alpha \geq 0 \wedge \beta \geq 0)} {(\alpha \geq 0)}} { \varone: \{ \alpha \in \typereal \}}.
      \end{equation}
    \end{example}
  \begin{example}
    We now look at the Rf rule, that deals with functions from 
    $\funsymbols$. Using this rule, we can show that:
     \begin{equation}
        \varone: \{ \alpha \in \typereal\}, \vartwo: \{\beta \in \typereal \}  {\imprr {(\alpha \geq 0 \wedge \beta \geq 0)} {(\gamma \geq 0)}} { \makereal{min}(x,y) : \{ \gamma \in \typereal \}}.
      \end{equation}
    \end{example}
Before giving the refined typing rule for the if-then-else construct, we also illustrate on an example how the rules in Figure~\ref{fig:refined_typing_system} allow us to exploit the continuity informations we have on  functions in $\funsymbols$, compositionally.
\begin{example}
  Let $f : \RR \to \RR$ be the function defined as: $f(x) = \begin{cases} -x \text{ if }x < 0 \\  x+1 \text{ otherwise}\end{cases}$. Observe that we can actually regard $f$
  as represented by the program in Figure~\ref{fig:non_continuous_program}---but we consider it as a primitive function in $\funsymbols$ for the time being, since we have not introduced the typing rule for the if-then-else construct, yet. Consider the program:
  $$\termone = \abs{\varone, \vartwo}{\makereal{f}(\makereal{min}(x,y))}. $$
  We see that $\sem \termone : \RR^2 \to \RR$ is continuous on the set 
  $\{(x,y) \mid x \geq 0 \wedge y \geq 0\}$, and that, moreover, 
  the image of $f$ on this set is contained on $[1, + \infty)$.  
  Using the rules in Figure~\ref{fig:refined_typing_system}, the fact that $f$ is 
  continuous on $\RR_{\geq 0}$, and that $min$ is continuous on $\RR^2$, 
  we see that our refined type system allows us to  prove $\termone$ to be continuous in the considered domain, i.e.:
  $$\impr \termone : \left(\{\alpha \in \typereal\}, \{\beta \in \typereal\} \right) \, \torr{(\alpha \geq 0 \wedge \beta \geq 0)}{(\gamma \geq 1)}\, \{ \gamma \in \typereal \}. $$
\end{example}

\subsection{Typing Conditionals}
We now look at the rule for the if-then-else construct: as can be seen
in the two programs in
Figure~\ref{fig:programs_with_branches}, the use of the if-then-else \emph{may} or \emph{may
not} induce discontinuity points. The crux here
is the behaviour of the two branches at the \emph{discontinuity points of
the guard function}. In the two programs represented in
Figure~\ref{fig:programs_with_branches}, we see that the only
discontinuity point of the guard is in $x=0$. However, in
Figure~\ref{fig:continuous_program} the two branches return the same value in $0$,
and the resulting program is thus continuous at $x=0$, while in
Figure~\ref{fig:non_continuous_program} the two branches do not
coincide in $0$, and the resulting program is discontinuous at $x=0$. We can
generalize this observation: for the program $\ifth \termone \termtwo
\termthree$ to be continuous, we need the branches $\termtwo$ and
$\termthree$ to be continuous respectively on the domain where
$\termone$ is $1$, and on the domain where $\termone$ is $0$, and
moreover we need $\termtwo$ and $\termthree$ to be continuous
\emph{and to coincide} on the points where $\termone$ is not
continuous. Similarly to the logical system designed by Chaudhuri et al~\cite{ChaudhuriGL10}, the coincidence of the branches in the discontinuity points is expressed as a set of logical rules by way of the \emph{observational equivalence.}  It should be observed that such an equivalence check is less problematic for first-order programs than it is for higher-order one (the authors of~\cite{ChaudhuriGL10} are able to actually check observational equivalence through an SMT solver).  On the other hand, various notions of equivalence which are included in contextual equivalence and sometimes coincide with it  (e.g., applicative bisimilarity, denotational semantics, or logical relations themselves) have been develloped for higher-order languages, and this starts to give rise to actual automatic tools for deciding contextual equivalence~\cite{DBLP:journals/pacmpl/Jaber20}.

We give in
Figure~\ref{figure:if-th-else} the typing rule for conditionals. The conclusion of the rule guarantees the continuity of the program $\ifth \termone \termtwo \termthree$ on a domain specified by a formula $\theta$. The premises of the rule ask for formulas $\theta_{\termfour}$ for $\termfour \in \{\termone, \termtwo, \termthree \}$ that specify continuity domains for the programs $\termone$, $\termtwo$, $\termthree$, and also for two additional formulas $\theta_{(\termone,0)}$ and $\theta_{(\termone,1)}$ that specify domains where the value of the guard $\termone$ is $0$ and $1$, respectively.  The target formula $\theta$, and the formulas $(\theta_{\termfour})_{\termfour \in \{\termone, \termtwo, \termthree, (\termone,1), (\termone,0)\}}$ are related by  two side-conditions. Side-condition~\eqref{side_cond_if_rule2} consists of the following four distinct requirements, that must hold for every point $a$ in the truth domain of $\theta$: \begin{enumerate*}[label=\roman*)]\item $a$ is in the truth domain of at least one of the two formulas $\theta_\termone$, $\theta_\termtwo$; \item if $a$ is not in $\theta_{(\termone,1)}$--(i.e.~we have no guarantee that $\termone$ will return $1$ at point $a$, meaning that the program $\termthree$ \emph{may} be executed)-- then $a$ must be in the continuity domain of $\termthree$;\item a condition symmetric to the previous one, replacing $1$ by $0$, and $\termthree$ by $\termtwo$; \item all points of possible discontinuity--(i.e.~the points $a$ such that $\theta^\termone$ does not hold)--must be in the continuity domain of both $\termtwo$ and $\termthree$, and as a consequence both $\theta^\termtwo$ and $\theta^\termthree$ must hold there  \end{enumerate*}.
   The side-condition~\eqref{side_cond_if_rule3} uses \emph{typed contextual equivalence
  $\equiv^{ctx}$} between terms to express that the two programs
$\termtwo$ and $\termthree$ must coincide on all
inputs such that $\theta_\termone$ does not hold--i.e.~that are not in the continuity domain of $\termone$. \RAPHAELLE{Observe that typed context equivalence here is defined with respect to the simple type system.} 

\begin{notation}
We use the following notations in Figure~\ref{figure:if-th-else}. When $\Gamma$ is a typing environement, we write $\gpart \Gamma$ and $\hpart \Gamma$ respectively the ground and higher-order part of $\Gamma$. Moreover, suppose we have a ground refined typing environment $\Theta = x_1:\{\alpha_1 \in \typereal\}, \ldots, x_n: \{\alpha_n \in \typereal\}$: we say that a logical assignment $\sigma$ is \emph{compatible with} $\Theta$ when $\{\alpha_i\mid 1\leq i \leq n\} \subseteq \support \sigma$. When it is the case, we build in a natural way the \emph{substitution associated to $\sigma$ along $\Theta$} by taking $\sigma^{\Theta}(x_i) = \makereal{\sigma(\alpha_i)}$.
  \end{notation}
\begin{center}
\begin{figure}[htbp]
  \fbox{
  \begin{minipage}{0.97 \textwidth}
{\center
$$
\AxiomC{$
  \begin{array}{c}
      \Gamma {\imprr {\theta_{\termone}}{ (\beta = 0 \vee \beta = 1)}} \termone: \{ \beta \in \typereal\}\\
       \Gamma {\imprr {\theta_{(\termone,0)}}{(\beta = 0)}} \termone: \{ \beta \in \typereal \}\\
        \Gamma {\imprr {\theta_{(\termone,1)}}{(\beta = 1)}} \termone: \{ \beta \in \typereal\}
      \end{array}$
}
\AxiomC{$\Gamma {\imprh{\theta_{\termtwo}(\eta)}} \termtwo:  T $}
\AxiomC{ $\Gamma {\imprh{\theta_{\termthree}(\eta)}} \termthree:  T $ }
\AxiomC{~\eqref{side_cond_if_rule2},~\eqref{side_cond_if_rule3} }
\LeftLabel{If}
\QuaternaryInfC{$ \Gamma {\imprh {\theta(\eta)}} \ifth \termone \termtwo \termthree : T $}
\DisplayProof
$$
    Where, by abuse of notation, $\psi(\eta)$ should be read as $\psi$ when $T$ is a higher-order type, and  as $\psi \leadsto \eta$ when $T$ is a ground type. The side-conditions~\eqref{side_cond_if_rule2}, ~\eqref{side_cond_if_rule3} are given as:
    \begin{varenumerate}
      \item\label{side_cond_if_rule2}  $\models \theta \Rightarrow  \left((\theta^{\termtwo} \vee \theta^{\termthree}) \wedge (\theta^{(\termone,1)} \vee \theta^{\termthree}) \wedge (\theta^{(\termone,0)} \vee \theta^{\termtwo}) \wedge (\theta_\termone \vee (\theta_\termtwo \wedge \theta_\termthree))\right)$.
      \item \label{side_cond_if_rule3}
 $ \text{For all logical assignment } \sigma \text{ compatible with }\gpart \Gamma, \sigma \models \theta \wedge \neg \theta_{\termone}$ implies $ \hpart \Gamma \imp  \termtwo{\sigma^{\gpart \Gamma}} \equiv^{ctx} \termthree {\sigma^{\gpart \Gamma}}.$
    \end{varenumerate}
}
  \end{minipage}}
    \caption{Typing Rules for the if-then-else construct}\label{figure:if-th-else}
 \end{figure}
\end{center}

 %
 \begin{example}
   Using our if-then-else typing rule, we can indeed type the program in Figure~\ref{fig:continuous_program} as expected, i.e.~:
   $$\imp \abs \varone {\ifth{x < 0}{1}{x+1}} : \{\alpha  \in \typereal  \} \torr{\top}{\top}\{ \beta \in \typereal\} .$$
   \end{example}

   \subsection{Open-logical Predicates for Refinement Types}
   Our goal in this section is to show the correctness of our refinement type systems, that we state below.
   \begin{theorem}\label{theorem:correctness_refinement_types}
     Let $\termone$ be any program such that:
     $$\varone_1:\{\alpha_1 \in \typereal\}, \ldots, \varone_n:\{ \alpha_n \in \typereal\} \imprr{\theta}{\theta'} \termone:\{\beta \in \typereal \}.$$
     Then it holds that:
     \begin{varitemize}
     \item $\sem \termone(\domain {\theta}^{\alpha_1, \ldots, \alpha_n}) \subseteq \domain {\theta'}^{\beta} $;
       \item $\sem \termone$ is sequentially continuous on $\domain {\theta}^{\alpha_1, \ldots, \alpha_n} $ .
       \end{varitemize}
   \end{theorem}
  
   As a first step, we show that our if-then-else rule is reasonnable, i.e. that it behaves well with primitive functions in $\funsymbols$. More precisely, if  we suppose that the functions $f,g_0, g_1$ are such that the premises of the if-then-else rule hold, then the program $\ifth{\makereal f(x_1, \ldots,x_n)}{\makereal {g_1}(x_1, \ldots,x_n))}{\makereal {g_0}(x_1, \ldots,x_n)}$ is indeed continuous in the domain specified by the conclusion of the rule. 
   This is precisely what we prove in the following lemma.
   \begin{lemma}\label{lemma:sound_math_condition_2}
      Let  $f,g_0,g_1 : \RR^n \to \RR$ be functions in $\funsymbols$, and $\Theta = x_1:\{\alpha_1 \in \typereal\}, \ldots, x_n:\{\alpha_n \in \typereal \}$. We denote $\vec \alpha$ the list of logical variables $\alpha_1, \ldots, \alpha_n$.  We consider logical formulas $\theta$ and $\theta_{f}, \theta_{(f,0)}, \theta_{(f,1)}, \phi_{g_0}, \phi_{g_1}$ that have their logical variables in $\vec \alpha$, and such that:
\begin{varenumerate}
\item\label{item:cond1} $f$ is continuous on $\domain \theta^{\vec \alpha}$ with $f(\domain {\theta_f}^{\vec \alpha})\subseteq \{0,1\}$ and $f(\domain{\theta_{(f,b)}}^{\vec \alpha}) \subseteq \{b\}$ for $b \in \{0,1\}$.
  \item\label{item:cond2} $g_0$ and $g_1$ are continuous on $\domain{\phi_{g_0}}^{\vec \alpha}$, and $\domain{\phi_{g_1}}^{\vec \alpha}$ respectively, and   $(\alpha_1 \mapsto a_1, \ldots,  \alpha_n \mapsto a_n )\models \theta \wedge \neg \theta_{f}$ implies $g_0(a_1, \ldots, a_n) = g_1(a_1, \ldots, a_n)$;
  \item\label{item:cond3} $\models  \theta \Rightarrow  \left((\phi_{g_1} \vee \phi_{g_0}) \wedge (\theta_{(f,0)} \vee \phi_{g_1}) \wedge (\theta_{(f,1)} \vee \phi_{g_0}) \wedge (\theta_f \vee (\phi_{g_0} \wedge \phi_{g_1}))\right).$ 
\end{varenumerate}
Then it holds that:
$$\sem{\forget \Theta \imp \ifth {\makereal f (x_1, \ldots,x_n)} {\makereal{g_1}(x_1, \ldots,x_n)}{\makereal{g_0}(x_1, \ldots,x_n)}: \typereal}$$
is continuous on $\domain \theta^{\vec \alpha}$.
   \end{lemma}
   \begin{proof}
\shortv{The proof can be found in the extended version \cite{EV}.}\longv{It is a direct consequence of Lemma~\ref{lemma:sound_math_condition_2} below. More precisely, we take $U = \domain{\theta}^{\vec \alpha}$, $U_b = \domain{\theta_{(f,b)}}^{\vec \alpha}$ for $b \in \{0,1\}$, and $L = \domain{\phi_{g_0}}^{\vec \alpha}$, $D = \domain{\phi_{g_1}}^{\vec \alpha}$. }
     \end{proof}
   \longv{
   \begin{lemma}\label{lemma:sound_math_condition_2}
  Let $X, L,D, U \subseteq \RR^n$, and $f,g_0,g_1 : \RR^n \to \RR$ be such that :
\begin{varenumerate}
\item\label{item:cond1} $f$ is continuous on $X$ with $f(X)\subseteq \{0,1\}$, $f(U_0) \subseteq \{0\}$, $f(U_1) \subseteq \{1\}$.
  \item\label{item:cond2} $g_0$ is continuous on $L$, and $g_1$ is continuous on $D$,  $g_0(x) = g_1(x)$ for every $x \in U \setminus X$;
  \item\label{item:cond3} $x \in U$ implies $x \in (L \cup D) \cap (L\cup  U_1) \cap (D \cup U_0) \cap (X \cup (L \cap D))$.
\end{varenumerate}
Then it holds that $\mathtt{cond}(f,g_1,g_0): x \in \RR^n \mapsto \begin{cases} g_1(x) \text{ if } f(x) = 1 \\ g_0(x) \text{otherwise}. \end{cases}$ is continuous on $U$.
\end{lemma}
\longv{\begin{proof}
  Let $x \in U$, and $(u_n)_{n \in \NN}$ a sequence of elements in $U$ such that $\lim_{n \to \infty } u_n = x$. We want to show that $\lim_{n \to \infty}\mathtt{cond}(f,g_1,g_0)(u_n) = \mathtt{cond}(f,g_1,g_0)(x)$.
   \begin{varitemize}
   \item We first suppose that $x \not \in X$. We split the sequence $(u_n)_{n \in \NN}$ in two sub-sequences $(v_n)_{n \in \NN}$, $(w_n)_{n \in \NN}$ such that for every $n \in \NN$, $f(v_n) \neq 1$ and $f(w_n) =1 $. Observe that at least one of those subsequences must be infinite. By condition~\ref{item:cond3}, we can see that all the $v_n$ are in $L$, and all the $w_n$ are in $D$. It means that whenever the sequence $(v_n)_{n \in \NN}$ is infinite, then $\lim \mathtt{cond}(f,g_1,g_0)(v_n) =  \lim g_0(v_n) = \lim g_0(x)$, and similarly whenever the sequence $(w_n)_{n \in \NN}$ is infinite then $\lim \mathtt{cond}(f,g_1,g_0)(w_n) = \lim g_1(w_n) = \lim g_1(x)$. From there, we conclude using~\ref{item:cond2} that tell us $g_0(x) = g_1(x)$.
   \item We now suppose that $x \in X$. Observe first that by hypothesis it holds that $f(x) \in \{0,1\}$. We split the sequence $(u_n)_{n \in \NN}$ in two sub-sequences $(v_n)_{n \in \NN}$, $(w_n)_{n \in \NN}$ such that the first one regroup all the $u_n$ that are in $X$, and the second one all the $u_n$ that are not in $X$.
     \begin{varitemize}
\item We first show that whenever $(v_n)_{n \in \NN}$ is infinite then $\lim \mathtt{cond}(f,g_1,g_0)(v_n) = \lim \mathtt{cond}(f,g_1,g_0)(x)$. We can for instance suppose that $f(x) = 0$--the reasoning is similar in the other case, i.e.~$f(x) = 1$--thus $x \in L$. Observe that since $f$ is continuous on $X$, $f(X) \subseteq \{0,1\}$, and both the $v_n$ and $x$ are in $X$, it holds that $\lim f(v_n) = f(x)$: it means that there exists $N \in \NN$, such that for every $n \geq N$, $f(v_n) = 0$, and as a consequence also $v_n \in L$. From there we can conclude--by continuity of $g_0$ on $L$--that:
  $$\lim \mathtt{cond}(f,g_1,g_0)(v_n) = \lim g_0(v_n) = g_0(x) =  \mathtt{cond}(f,g_1,g_0)(x) .$$
\item We now show that whenever $(w_n)_{n \in \NN}$ is infinite then $\lim \mathtt{cond}(f,g_1,g_0)(w_n) = \lim \mathtt{cond}(f,g_1,g_0)(x)$. Since the $w_n$ are in $U \setminus X$, by~\ref{item:cond2} it holds that $g_0(w_n) = g_1(w_n)$. Let us for instance suppose that $f(x) = 0$--the reasoning would be similar if $f(x) = 1$. It means that $x \in L$, and since moreover $U \setminus X \subseteq L \cap D$, the $w_n$ also are in $L$. As a consequence, we can use the fact that $g_1$ is continuous on $L$, and we obtain:
   $$\lim \mathtt{cond}(f,g_0,g_1)(w_n) = \lim g_1(w_n) = g_1(x) =  \mathtt{cond}(f,g_0,g_1)(x) .$$
  \end{varitemize}
 \end{varitemize}
\end{proof}}}
   Similarly to what we did in Section~\ref{section:conservation-theorem}, we are going to show Theorem~\ref{theorem:correctness_refinement_types} by way of a logical predicate. Recall that the logical predicate we defined in Section~\ref{section:conservation-theorem} consists actually of \emph{three} kind of predicates---all defined in Definition~\ref{def:open-logical-predicate} of Section~\ref{section:conservation-theorem}: 
   $\funpred^{\envreal}_{\typeone}$,  $\funpred^{\envreal}_{\Gamma}$, 
   $\funpred^{\envreal, \Gamma}_{\typeone}$, where $\envreal$ ranges over ground typing environments, $\Gamma$ ranges over arbitrary environments, and $\typeone$ is a type. 
   The first predicate $\funpred^{\envreal}_{\typeone}$ contains  admissible terms $\termone$ of type $\envreal \imp \termone : \typeone$, the second predicate $\funpred^{\envreal}_{\Gamma}$ contains  admissible substitutions $\gamma$ that associate to every $(\varone: \typeone)$ in $\Gamma$ a term of type $\typeone$ under the typing context $\envreal$, and the third predicate $\funpred^{\envreal, \Gamma}_{\typeone}$ contains admissible terms $\termone$ of type $\Gamma, \Theta \imp \termone: \typeone$.
   
   Here, we need to adapt all three kinds of logical predicates to a 
   refinement scenario: first, we replace $\typeone$ and $\Theta$, $\Gamma$ with refinement types and refined typing contexts respectively. Moreover, for technical reasons, we also need to \emph{generalize} our typing contexts, by allowing them to be annotated 
   with any subset of $\RR^n$ instead  of restricting ourselves to those subsets generated by logical formulas. Due to this further complexity, we split our definition of logical predicates in two: we first define the counterpart of the ground typing context predicate $\funpred^{\envreal}_{\typeone}$ in Definition~\ref{def:red_candidates_ground_contexts}, then the counterpart of the predicate for substitutions $\funpred^{\envreal}_{\Gamma}$  and the counterpart of the predicates $\funpred^{\envreal, \Gamma}_{\typeone} $ for higher-order typing environment in Definition~\ref{ldef:adm_substs_ref_types}.

 Let us first see how we can adapt the predicates $\funpred^{\envreal}_{\typeone}$ to our refinement types setting. Recall that in Section~\ref{section:conservation-theorem}, 
 we defined the predicate $\funpred^{\envreal}_{\typereal}$ as the collection 
of terms $\termone$ such that $\envreal \imp t:\typereal$, and its semantics 
$\sem{\envreal \imp t:\typereal}$ 
belongs to $\funsymbols$. As we are interested in local continuity properties, we need to 
build a predicate expressing local continuity constraints. 
Moreover, in order to be consistent with our two arrow constructs and our two kinds of typing judgments, we actually need to consider also \emph{two} kinds of logical 
predicates, depending on whether the target type we consider is a real type or 
an higher-order type. We thus introduce the following logical predicates:
 $$\redgr{\Theta}{X}{\phi}{F}; \qquad \redgh{\Theta}{X}{H};$$
 where $\Theta$ is a  ground typing environment, $X$ is a subset of $\RR^n$, $\phi$ is a logical formula, and, as usual, $F$ ranges over the real refinements types, while $H$ ranges over the higher-order refinement types.  As expected, $X$ and $\phi$ are needed to encode continuity constraints inside our logical predicates.
 
   \begin{definition}\label{def:red_candidates_ground_contexts}
   Let $\Theta$ be a \emph{ground}  typing context of length $n$, $F$ and $H$ refined ground type and higher-order type, respectively.
   We define families of predicates on terms 
   $\redgr{ \Theta}{Y}{\phi}{F}$ and $\redgh{\Theta}{Y}{H}$, 
   with $Y \subseteq \RR^n$ and $\phi$ a logical formula, as specified in 
   Figure~\ref{figure:open_logical_predicate_refinement_types}.
  \end{definition}

   \begin{center}
\begin{figure}[htbp]
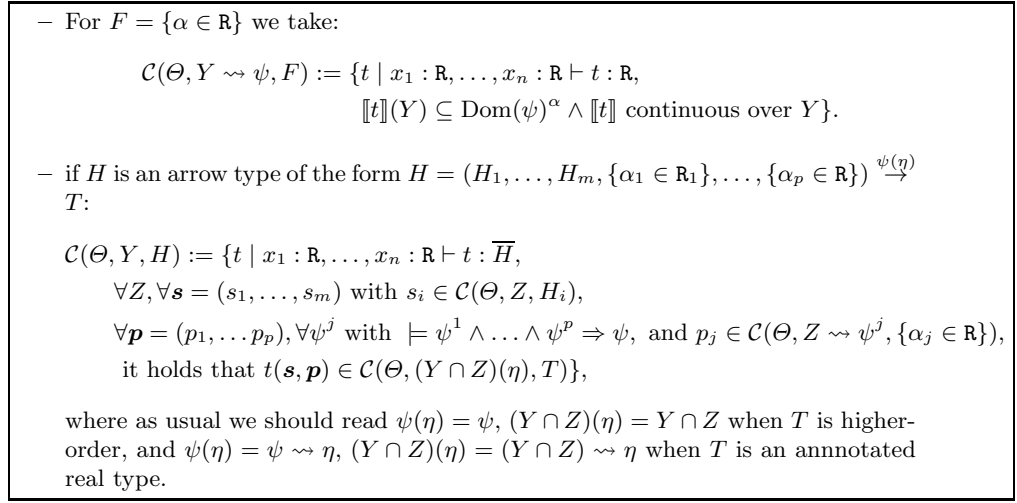

  \fbox{
    \begin{minipage}{.97\textwidth}
\begin{varitemize}
  \item For $F = \{ \alpha \in \typereal\}$  we take:
\begin{align*}
   \redgr{ \Theta}{Y}{\psi}{F}
    := \{ &\termone \mid  \varone_1:\typereal, \ldots, \varone_n : \typereal \imp \termone : \typereal, \\ & 
      \sem{\termone}(Y)\subseteq \domain \psi^\alpha \wedge \sem {\termone} \text{ continuous over } Y \}.
    \end{align*}
\item if $H$ is an arrow type of the form $H = (H_1, \ldots, H_m,  \{\alpha_1 \in \typereal_1 \}, \ldots, \{\alpha_p \in \typereal\}) \tor {\psi(\eta)} {T}$:
\begin{align*}
    & \redgh{\Theta}{Y}{H}:=  \{ \termone \mid  \varone_1:\typereal, \ldots, \varone_n : \typereal \imp \termone :  \forget{H}, \\
  & \qquad  \forall Z, \forall \vec \termtwo = (\termtwo_1, \ldots, \termtwo_m)  \text{ with } \termtwo_i \in  \redgh{ \Theta}{Z}{H_i}, \\
  & \qquad \forall \vec \termthree = (\termthree_1, \ldots \termthree_p), \forall \psi^j \text{ with } \models \psi^1 \wedge \ldots \wedge \psi^p \Rightarrow \psi, \text{ and } \termthree_j \in  \redgr{\Theta}{Z}{\psi^j}{\{\alpha_j \in \typereal \}} , \\
  & \qquad \text{ it holds that } \termone (\vec \termtwo, \vec \termthree) \in \redgh{\Theta}{(Y\cap Z)(\eta)}{T} \},
\end{align*}
where as usual we should read $\psi(\eta) = \psi$, $(Y \cap Z)(\eta)  = Y \cap Z$ when $T$ is higher-order, and $\psi(\eta) = \psi \leadsto \eta$, $(Y \cap Z)(\eta) =(Y \cap  Z) \leadsto \eta$ when $T$ is an annnotated real type.
 \end{varitemize}
\end{minipage}}
  \caption{Open logical predicate for refinement types.}\label{figure:open_logical_predicate_refinement_types}
\end{figure} 
\end{center}

 \begin{example}\label{ex:logical_predicate_ref_types_1}
  We illustrate Definition~\ref{def:red_candidates_ground_contexts} on some examples. We denote by
  $B^{\circ}$ the open unit ball in $\RR^2$, 
  i.e.~$B^{\circ} =\{(a,b) \in \RR^2 \mid a^2 + b^2 < 1\}$. 
  We consider the ground typing context $\Theta = x_1:\{ \alpha_1 \in \typereal\}, x_2:\{\alpha_2 \in \typereal\}$.
\begin{varitemize}
  \item  We look first at the predicate $\redgr{\Theta}{B^{\circ}}{(\beta > 0)}{\{ \beta \in \typereal \}}.$ It consists of all programs $x:\typereal \imp t:\typereal$ such that $\sem \termone$  is continuous on the open unit ball, and takes only strictly positive values there.
\item We look now at an example when the target type $T$ is \emph{higher-order}. 
  We take  $H = \{\beta_1 \in \typereal\} \torr{(\beta_1 \geq 0)}{(\beta_2 \geq 0)}{\{\beta_2 \in \typereal\}}$, and we look at the logical predicate $\redgh{\Theta}{B^{\circ}} H.$
    We are going to show that the latter contains, for instance, 
    the program: $$\termone = \abs w \makereal{f}(w,x_1^2 + y_1^2) \qquad \text{ where }f(w,a) =  \frac w {1-a} \text{ if } a < 1; 0 \text{ otherwise.}$$
    Looking at Figure~\ref{figure:open_logical_predicate_refinement_types}, we see that it is enough to check that for any $Y \subseteq \RR^2$ and any $\termtwo \in \redgr{\Theta}{Y}{(\beta_1 \geq 0)}{\{\beta_1 \in \typereal\}}$, it holds that: $$\termone \termtwo \in  \redgr{\Theta}{B^\circ \cap Y}{(\beta_2 \geq 0)}{\{\beta_2 \in \typereal\}}.$$
    \end{varitemize}
 \end{example}
Our overall 
goal---in order to prove Theorem~\ref{theorem:correctness_refinement_types}---is to show the counterpart of 
Fundamental Lemma of Section~\ref{section:conservation-theorem} (i.e. 
Lemma~\ref{lemma:fundamental-lemma-logical-predicate}), 
which states that the logical predicate $\funpred^{\envreal}_{\typereal}$ contains all well-typed terms. 
This lemma talks only about the logical predicates for 
\emph{ground typing contexts}, so we can state it as of now, 
but its proof is based on the fact that we dispose of 
the \emph{three} kinds of logical predicates.  
Observe that from there, Theorem~\ref{theorem:correctness_refinement_types}
follows just from  the definition of the logical predicates on base types.
\begin{proposition}\label{proposition:well-typed-programs_in_pred_reftypes}
  Let $\termone$ be a program such that
  $\Theta \imprr{\psi}{\psi'} \termone:\{\beta \in \typereal\}.$ Then $\termone$ is in the predicate $\redgr{\Theta}{\psi}{\psi'}{\{\beta \in \typereal\}}.$
  \end{proposition}

Similarly to what we did for 
Lemma~\ref{lemma:fundamental-lemma-logical-predicate} 
in Section~\ref{section:conservation-theorem},
we need first to define the logical predicates for \emph{substitutions} and \emph{higher-order typing contexts}.
We do this in Definition~\ref{ldef:adm_substs_ref_types} below. 
As before, they consist in an adaptation to our refinement types framework of the open logical predicates  $\funpred^{\envone}_{\envreal}$ and  
$\funpred^{\envreal,\Gamma}_{\typeone}$  of Section~\ref{section:conservation-theorem}: 
as usual, we need to add continuity annotations, and distinguish whether the target type is a ground type or an higher-order type.
\RAPHAELLE{
  \begin{notation}
    We need to first introduce the following notation: let $\Gamma$, $\Theta$ be two ground non-refined typing environments of length $m$ and $n$ respectively--and with disjoint support. Let $\gamma : \support \Gamma \to \{ \termone \mid \forget \Theta \imp \termone: \typereal\}$ be a substitution. We write $\sem \gamma$ for the real-valued function:
     \begin{align*}
    \sem \gamma : &\RR^n \to \RR^{n+m} \\
    &\vec a  \mapsto (\vec a, \sem{\gamma(x_1)}{(\vec a)}, \ldots, \sem{\gamma(x_m)}{(\vec a)})
     \end{align*}
    \end{notation}
  \begin{definition}\label{ldef:adm_substs_ref_types}
  Let $\Theta$ be a  ground typing environment of length $n$, and $\Gamma$ an arbitrary typing environment. We note $n$ and $m$ the lengths of respectively $\Theta$ and $\gpart \Gamma$.
  \begin{varitemize}
    \item Let $Z \subseteq \RR^n, W \subseteq \RR^{n+m}$. We define $\redgr \Theta{Z}{W} \Gamma$ as the set of those substitutions $\gamma : \support \Gamma \to \{ \termone \mid \forget \Theta \imp \termone: \typereal\}$ such that:
  \begin{varitemize}
  \item $\forall (\varone: H) \in {\hpart \Gamma}$, $\gamma(\varone) \in \redgh{\Theta}{Z}{H}$,
  \item  $\sem {\gamma_{\mid \gpart \Gamma}}: \RR^n \to \RR^{n+m}$ sends continuously $Z$ into $W$;
  \end{varitemize}
\item Let $W \subseteq \RR^{n+m}$, $F = \{\alpha \in \typereal\}$ an annotated real type, and $\psi$ a logical  formula with $\vars \psi \subseteq \{\alpha\}$. We define:
    \begin{align*}
& \redgr{(\Gamma; \Theta)}{W}{\psi}{F} 
      := \{ \termone \mid \forget {\Gamma, \Theta} \imp \termone : \typereal \\
&\qquad \wedge \forall X \subseteq \RR^n,  \forall \gamma \in \redgr{\Theta}{X}{W}{\Gamma},\, \termone \gamma \in \redgr{\Theta }{X}{\psi}{F}\}.
    \end{align*}
    \item Let $W \subseteq \RR^{n+m}$, and $H$ an higher-order refined  type. We define :
    \begin{align*}
&\redgh{(\Gamma; \Theta)}{W}{H} 
:= \{ \termone \mid \forget {\Gamma, \Theta} \imp \termone : \forget H \\
&\qquad \wedge {\forall X \subseteq \RR^n},  \forall \gamma \in \redgr{\Theta}{X}{W}{\Gamma}.\ \termone \gamma \in \redgh{{\Theta}}{X }{H}\}.
\end{align*}
\end{varitemize}
  \end{definition}}
  \begin{example}
    We illustrate Definition~\ref{ldef:adm_substs_ref_types} on an example. We consider the same context $\Theta$ as in Example~\ref{ex:logical_predicate_ref_types_1}, i.e. $\Theta = x_1:\{ \alpha_1 \in\typereal\}, x_2 : \{ \alpha_2 \in \typereal\}$, and we take $\Gamma = x_3: \{ \alpha_3 \in \typereal\}, z:H$, with  $ H =\{\beta_1 \in \typereal \} \torr{(\beta_1 \geq 0)}{(\beta_2 \geq 0)}{\{\beta_2 \in \typereal\}}$.
      We are interested in the following logical predicate for substitution:
    $$\redgr \Theta{B^{\circ}}{\{ (v,\lvert v \rvert ) \mid v \in B^{\circ})\}}{\Gamma}$$ where the norm of the couple $(a,b)$ is taken as: $  \lvert (a,b) \rvert = \sqrt {a^2 + b^2}$.
      We are going to build a substitution $\gamma :\{x_3,z\} \to \STLambdarealfun$ that belongs to this set. We take:
    \begin{varitemize}
    \item $\gamma(z) = \abs w \makereal{f}(w,x_1^2 + x_2^2)$ where $f(w,a) =  \frac w {1-a} \text{ if } a < 1; 0$ otherwise. 
    \item $\gamma(x_3) = \makereal{(\sqrt{\cdot})} (x_1^2 + x_2^2)$. 
    \end{varitemize}
    We can check that the requirements of Definition~\ref{ldef:adm_substs_ref_types} indeed hold for $\gamma$, i.e.~that:
    \begin{varitemize}
    \item $\gamma(z) \in \redgh{\Theta}{B^\circ}{H}$---see Example~\ref{ex:logical_predicate_ref_types_1};
      \item $\sem {\gamma_{\mid \gpart \Gamma}} : \RR \times \RR \to \RR^3$ is continuous on $B^\circ$, and moreover sends  $B^{\circ}$ into $\{ (v,\lvert v \rvert ) \mid v \in B^{\circ})\}$. Looking at our definition of the semantics of a substitution, we see that $\sem {\gamma_{\mid \gpart \Gamma}} (a,b)= (a,b,\lvert(a,b) \rvert)$, thus the requirements above hold.
      \end{varitemize}
    \end{example}
  
%
  
  Now that we have defined logical predicates for refinement types, we proceed towards the proof of Proposition~\ref{proposition:well-typed-programs_in_pred_reftypes}, by following the road-map defined in Section~\ref{section:conservation-theorem}. The next step at this point is to show that our predicates for ground typing environments are closed under denotational semantics, i.e.~that if some term $\termone$ is in some logical predicate  $\redgr{\Theta}{X}{\psi} F$ or  $\redgh{\Theta}{X} H$, then any term $\termtwo$ with the same denotational semantics as $\termone$ is also in this predicate: 
  this is the way one should read Lemma~\ref{lemma:preservation_denot_sem_ref_types} below. Recall that a similar result was a key component in the proof of the Fundamental Lemma---Lemma~\ref{lemma:fundamental-lemma-logical-predicate}---in Section~\ref{section:conservation-theorem}.
 \begin{lemma}\label{lemma:preservation_denot_sem_ref_types}
\begin{align*}
  & \termone \in \redgr{\Theta}{X}{\psi} F \, \wedge\, \sem{\forget \Theta \imp \termone:\forget F} = \sem{\forget \Theta \imp \termtwo: \forget F} \Longrightarrow \termtwo \in \redgr{\Theta}{X}{\psi} F \\
   & \termone \in \redgh{\Theta}{X} H \, \wedge\, \sem{\forget \Theta \imp \termone:\forget H} = \sem{\forget \Theta \imp \termtwo: \forget H} \Longrightarrow \termtwo \in \redgh{\Theta}{X} H.
    \end{align*}
  \end{lemma}
  \longv{
    \begin{proof}
The proof is by induction on the target type ($F$ or $H$).
      \end{proof}
  }

 At this point, we need to depart from Section~\ref{section:conservation-theorem}, and to show an additional lemma, that talks specifically about the way our continuity annotations behave when plugged into logical predicates. Indeed, recall that in the refined typing system we used  side conditions of the form $\models \psi \Rightarrow \phi$. In order to show Proposition~\ref{proposition:well-typed-programs_in_pred_reftypes}, we need to look at how this kind of conditions is reflected into our logical predicates. 
 This is what Lemma~\ref{lemma:monotony_logical_predicates} states.
 \begin{lemma}\label{lemma:monotony_logical_predicates}
  Let be $\Theta$ a ground typing environment, $\Gamma$ an arbitrary environment, and $H,F$ respectively an higher-order type and a ground type. We order the set of the subset of $\RR^n$ by inclusion, and the set of formulas by: $\psi \leq \psi'$ when $\models \psi \Rightarrow \psi'$. Then:
  \begin{varitemize}
  \item the maps $X, \psi \mapsto \redgr{\Theta}{X}{\psi} F$ and $X \mapsto \redgh{\Theta}{X} H$ are non-increasing in $X$ and non-decreasing in $\psi$.
    \item  the map $Z,W \mapsto \redgr \Theta{Z}{W} \Gamma$ is non-increasing in $Z$ and non-decreasing in $W$.
    \item the maps $X, \psi \mapsto \redgr{\Gamma;\Theta}{X}{\psi} F$ and $X \mapsto \redgh{\Gamma;\Theta}{X} H$ are non-increasing in $X$ and non-decreasing in $\psi$.
    \end{varitemize}
 \end{lemma}
 \longv{
   \begin{proof}
     Let $\Theta$ be a ground typing environment of size $n$, $\Gamma$ an arbitrary environment such that its ground part $\gpart \Gamma$ has length $m$.
     We first show by induction on the target type that  $X, \psi \mapsto \redgr{\Theta}{X}{\psi} F$ and $X \mapsto \redgh{\Theta}{X} H$ are non-increasing in $X$ and non-decreasing in $\psi$.
      \begin{varitemize}
      \item Let $F = \{ \alpha \mid \alpha \in \typereal\}$. Let $X \subseteq X'$, and $\psi \leq \psi'$. We suppose a term $\termone \in \redgr{\Theta}{X'}{\psi} F$. Our goal is to show that $\termone \in \redgr{\Theta}{X}{\psi'} F$. By hypothesis, we know that $\sem \termone$ is continuous on $X'$, and sends it into $\domain{\psi}^{\alpha}$. Since $X \subseteq X'$, $\sem \termone$ is also continuous on $X$, and moreover $\sem \termone (X) \subseteq \sem \termone (X') \subseteq \domain{\psi}^{\alpha}$. Since moreover $\psi \leq \psi'$, it also holds that $\domain \psi^{\alpha} \subseteq \domain {\psi'}^{\alpha}$. Summing up, we have shown that $\sem \termone$sends continuously $X$ into $\domain {\psi'}^\alpha$, thus we can conclude that $\termone \in \redgr{\Theta}{X}{\psi'} F$.
        \item We now suppose that $H = (T_1, \ldots, T_m) \tor{\phi(\eta)} T$. Let $X \subseteq X'$ and $\termone \in \redgh{\Theta}{X'} H$. Our goal is to show that $\termone \in \redgh{\Theta}{X} H$. Let $Z \subseteq \RR^n$, and formulas $(\phi_j)$ such that $\phi_1 \wedge \ldots \wedge \phi_m \Rightarrow \phi$, and  $\vec \termtwo = \termtwo_1, \ldots, \termtwo_m$ a sequence of terms such that $\termtwo_i \in \redgh{\Theta}{Z(\phi_i)}{T_i}$. We know by hypothesis that $\termone (\vec \termtwo) \in \redgh{\Theta}{X' \cap Z(\eta)}{T}$. Our goal is to show that also  $\termone (\vec \termtwo) \in \redgh{\Theta}{X \cap Z(\eta)}{T}$. To see this, it is enogh to apply our induction hypothesis to the target type $T$, using the fact that $X \cap Z \subseteq X' \cap Z$.
      \end{varitemize}
      We now show that the map $Z,W \mapsto \redgr \Theta{Z}{W} \Gamma$ is non-increasing in $Z$ and non-decreasing in $W$. Let $Z \subseteq Z' \subseteq \RR^n$,  $W \subseteq W' \subseteq \RR^{n+m}$ and $\gamma$ a subsitution such that $\gamma \in \redgr \Theta {Z'} W \Gamma$: our goal here is to show that $\gamma \in \redgr \Theta{Z}{W'} \Gamma$. We split the proof of this fact in two parts--the first one talks about higher-order variables in $\Gamma$, while the second one talks about real variables.
      \begin{varitemize}
      \item Let $(\varone:H) \in \hpart \Gamma$. We want to show that $\gamma(\varone) \in \redgh{\Theta}{Z}{H}$. By hypothesis--since $\gamma \in \redgr \Theta{Z'}{W} \Gamma$, it holds that $\gamma(\varone) \in \redgh{\Theta}{Z'}{H}$. From there, it is enough to apply the first point of Lemma~\ref{lemma:monotony_logical_predicates}--that we have proved above-- and we can conclude using the fact that $Z \subseteq Z'$.
        \item We need now to show that $\sem \gamma: \RR^n \to \RR^{n+m}$ is continuous on $Z$, and $\sem \gamma(Z) \subseteq W'$. To see this, it is enough to use the fact that by hypothesis, $Z \subseteq Z'$ and $W \subseteq W'$.
      \end{varitemize}
      We finally show that the maps $X, \psi \mapsto \redgr{\Gamma;\Theta}{X}{\psi} F$ and $X \mapsto \redgh{\Gamma;\Theta}{X} H$ are non-increasing in $X$ and non-decreasing in $\psi$. Let $F = \{\alpha \mid \alpha \in \typereal\}$, $X \subseteq X'$, and $\psi \leq \psi'$. We suppose that $\termone \in \redgh{\Gamma;\Theta}{X'}{H}$, and $\termtwo \in \redgr{\Gamma;\Theta}{X'}{\psi} F $: our goal is to show that also $\termone \in \redgh{\Gamma;\Theta}{X}{H}$ and $\termtwo \in \redgr{\Gamma;\Theta}{X}{\psi'} F $. To do that, we need to suppose $W \subseteq \RR^{n}$, and  $\gamma \in \redgr{\Theta}{W}{X}{\Gamma}$, and to show that $\termtwo \gamma \in \redgr{\Theta}{W}{\psi'}{F}$ and $\termone \gamma \in \redgh{\Theta}{W}{H}$. First, by the second point of the present lemma-that we have already shown before--we can see that $\gamma \in \redgr{\Theta}{W}{X'}{\Gamma}$. From there, we can apply the hypoethesis on $\termone, \termtwo$, and we obtain that $\termone \gamma \in \redgr{\Theta}{W}{\psi}{F}$ and $\termtwo \gamma \in \redgh{\Theta}{W}{H}$. To conclude from there, it is enough to apply the first point of the present lemma, using the fact that $\psi \leq \psi'$.
     \end{proof}
   }
We are now ready to do the last step of the proof of Proposition~\ref{proposition:well-typed-programs_in_pred_reftypes}, i.e.~to show the counterpart of the fundamental lemma of Section~\ref{section:conservation-theorem}. This fundamental lemma is essentially 
a stronger variant of 
Proposition~\ref{proposition:well-typed-programs_in_pred_reftypes}---in the sense where it talks not only about terms $\termone$ such that $\Theta \imp \termone: T$, but also of terms typable along \emph{non-ground typing environments}, i.e.~ $\Gamma \imp \termone: T$: it means that it implies Proposition~\ref{proposition:well-typed-programs_in_pred_reftypes} as a corollary. Observe that to state---and to show---this lemma, we need to use our 
\emph{three kinds} of logical predicates.

 
  \begin{lemma}[Fundamental Lemma]\label{lemma:fundamental_lemma_ref_types}
    Let $\Theta$ be a  ground typing context, and $\Gamma$ an arbitrary typing context--thus $\Gamma$ can contain both ground type variables and non-ground type variables.
    \begin{varitemize}
    \item  Suppose that $\Gamma, \Theta \imprr{\theta}{\eta} \termone: F$:
      then $\termone \in \redgr{\Gamma; \Theta}{\domain{\theta}}{\eta}{F}.$
    \item  Suppose that $\Gamma, \Theta \imprh{\theta} \termone: H$: then $\termone \in \redgh{\Gamma; \Theta}{\domain{\theta}}{H}.$
    \end{varitemize}
    \end{lemma}
    \shortv{ \begin{proof}
     The proof is by induction on the derivation of the refined typing judgment, and uses both Lemma~\ref{lemma:preservation_denot_sem_ref_types} and Lemma~\ref{lemma:monotony_logical_predicates}. The details can be found in the extended version~\cite{EV}.\end{proof}}
      \longv{
        Before proving Lemma~\ref{lemma:fundamental_lemma_ref_types}, we need to show an auxilliary lemma that will allow us to deal with the if-then-else case.
\begin{lemma}\label{lemma:if-then-else_reduction}
  Let $\vec{q}$ a sequence of terms products, i.e. of the form: $\vec q = (\termthree^1_1 \cc \termthree^1_{l_1}), \ldots,  (\termthree^k_1 \cc \termthree^k_{l_k}) $. Let $\Theta = (\varone_i: \{ \alpha_i \in \typereal\})_{1 \leq i \leq n}$ be a ground typing context, $\termone, \termtwo, \termthree$ such that $\forget{\Theta} \imp (\ifth \termone \termtwo \termthree){\vec q}: \forget T$.
  We  suppose that there exist $X \subseteq \RR^n$, and a family $ X_w \subseteq \RR^n$ with $w \in \{\termthree, \termtwo, \termone, (\termone,0),(\termone,1)\}$ that are compatible with $\Theta$ such that:
\begin{varitemize}
\item $\termtwo{\vec q} \in \redgh{ \Theta}{X_{\termtwo}(\eta)}{T};\, \termthree{\vec q} \in \redgh{ \Theta}{ X_{\termthree}(\eta)}{T};$
  \item $\termone \in \redgr{\Theta}{ X_{\termone}}{(\beta = 0 \vee \beta = 1)}{ \{\beta \in \typereal\}}$;
  \item $ \termone \in \redgr{\Theta}{X_{(\termone,0)}}{(\beta = 0)}{ \{\beta \in \typereal\}} \cap \redgr{\ \Theta}{X_{(\termone,0)}}{(\beta = 1)}{ \{\beta \in \typereal\}}$
    \end{varitemize}
and moreover the following additional conditions hold:
\begin{varenumerate}
   \item\label{cond2:if-then-else_reduction} $ X \subseteq  \left(( X_{\termtwo} \cup X_{\termthree}) \cap (X_{\termone,0} \cup X_{\termthree}) \cap (X_{\termone,1} \cup X_{\termtwo}) \cap (X_\termone \cup (X_\termtwo \cap X_\termthree))\right)$;
     \item\label{cond3:if-then-else_reduction}  $ \forall a = (a_1, \ldots,a_n) \in X$, $a \in X \setminus X_{\termone}$ implies:  
  $$ \imp \subst{\subst {\termtwo \vec q} {x_1}{a_1} \ldots}{x_n}{a_n} \equiv^{ctx} \subst{\subst {\termthree \vec q} {x_1}{a_1} \ldots}{x_n}{a_n}.$$
\end{varenumerate}
  Then it holds that also $(\ifth \termone \termtwo \termthree) \vec q \in \redgh{\Theta}{X(\eta)}{T}.$
\end{lemma}
\begin{proof}
  The proof is by induction on $T$-- and uses Lemma~\ref{lemma:sound_math_condition_2}.
  \begin{varitemize}
  \item We first suppose that $T = \{\alpha \in \typereal \}$. It means that we can reformulate our goal as:
    \begin{equation}
\sem{(\ifth \termone \termtwo \termthree) \vec q} \text{ sends continuously } X \text{ into }\domain{\eta}^{\alpha}
\end{equation}
\item We now suppose that $T = (H_1, \ldots, H_m, \{\alpha_1 \in \typereal\}, \ldots, \{\alpha_n \in \typereal\} \tor{\psi(\phi)} T')$. We suppose  $ \vec \termfour = (\termfour_1, \ldots, \termfour_m)$ ,  $Y^i$ with $\termfour_i \in  \redgh{ \Theta}{Y^i}{H_i}$, $\vec \termfour' = (\termfour'_1, \ldots \termfour'_p)$,$Y^j,  \psi^j$ with $\models \psi^1 \wedge \ldots \wedge \psi^p \Rightarrow \psi$,  and $\termfour'_j \in  \redgr{\Theta}{Y^j}{\psi^j}{\{\alpha_j \in \typereal \}}$. We denote $Y = \wedge_{1 \leq i \leq n} Y_i \wedge_{1 \leq j \leq p} Y_j$. Our objective is to show:
  \begin{equation}
(\ifth \termone \termtwo \termthree) \vec q (\vec \termfour, \vec{\termfour'}) \in  \redgh{\Theta}{X \wedge Y (\phi)}{T'}
\end{equation}
We first use our hypothesis on $\termtwo{\vec q}$, $\termthree{\vec q}$\ldots. We obtain that:
\begin{varitemize}
\item $\termtwo{\vec q}{(\vec \termfour, {\vec \termfour'})} \in \redgh{\Theta}{X_{\termtwo} \wedge Y(\phi)}{T'}$;
  \item $\termthree{\vec q}{(\vec \termfour, {\vec \termfour'})} \in \redgh{\Theta}{X_{\termthree} \wedge Y (\phi)}{T'}$.
  \end{varitemize}
  From there, we would like to apply the induction hypothesis to the type $T'$. To do that, it is enough to check that:
  \begin{varitemize}
   \item $ X \wedge Y \subseteq  \left(( X_{\termtwo} \wedge Y \cup X_{\termthree} \wedge Y) \cap (X_{\termone,0} \cup X_{\termthree} \wedge Y) \cap (X_{\termone,1} \cup X_{\termtwo} \wedge Y) \cap (X_\termone \cup (X_\termtwo \wedge Y \cap X_\termthree \wedge Y))\right)$; We can see that it is indeed the case by usin Condition~\ref{cond2:if-then-else_reduction} in our hypothesis.
     \item  $ \forall a = (a_1, \ldots,a_n) \in X \wedge Y$, $a \in X \wedge Y \setminus X_{\termone}$ implies:  
  $$ \imp \subst{\subst {\termtwo \vec q(\vec \termfour, \vec \termfour'')} {x_1}{a_1} \ldots}{x_n}{a_n} \equiv^{ctx} \subst{\subst {\termthree \vec q(\vec \termfour, \vec{\termfour'}} {x_1}{a_1} \ldots}{x_n}{a_n}.$$ We see that it is indeed the case becauseof Condition~\ref{cond3:if-then-else_reduction} in the hypothesis, and the definition of context equivalence.
\end{varitemize}
\end{varitemize}
  \end{proof}
        We are now finally ready to do the proof of Lemma~\ref{lemma:fundamental_lemma_ref_types}--our fundamental lemma for showing correctness of the refinement type system, that we state again below.
         \begin{lemma}[Fundamental Lemma]
    Let $\Theta$ be a  ground typing context, and $\Gamma$ an arbitrary typing context--thus $\Gamma$ can contain both ground type variables and non-ground type variables.
    \begin{varitemize}
    \item  We suppose that $\Gamma, \Theta \imprr{\theta}{\eta} \termone: F$.
      Then $\termone \in \redgr{\Gamma; \Theta}{\domain{\theta}}{\eta}{F}.$
    \item  We suppose that $\Gamma, \Theta \imprh{\theta} \termone: H$.
      Then $\termone \in \redgh{\Gamma; \Theta}{\domain{\theta}}{H}.$
    \end{varitemize}
    \end{lemma}
        \begin{proof}
  The proof is by induction on the derivation of the refined typing judgment. Observe that depending of the target type, our typing judgment can be of the form $\Gamma, \Theta \imprh{\theta} \termone: H$ or $\Gamma, \Theta \imprr{\theta}{\eta} \termone: F$ .
    Since the proof path is similar, we are going to treat these two cases \emph{in the same time}: as an abuse of notation, we will note as usual $\theta(\eta)$ to means that we need to read $\theta$ when we consider the case when the target type is a $H$, and $\theta \leadsto \eta$ when it is a $F$.
    \begin{varitemize}
    \item We first suppose that $\termone = \varone$, such that $\varone$ is an higher-order variable in $\Gamma$--i.e.~$(x:H) \in \Gamma$. Then the typing rule used is as follows--where $\Gamma'$ denotes the environment $\Gamma$ minus $(\varone:H)$:
      {\scriptsize $$\AxiomC{}
         \LeftLabel{var-H}
        \UnaryInfC{$ \Gamma', x: H, \Theta {\imprh \psi} x: H $}
        \DisplayProof$$
      }
      We suppose $X \subseteq \RR^n$, and $\gamma$ a substitution such that $\gamma \in  \redgr{\Theta}{X}{\domain \theta}{\Gamma} $. Our goal is to show that:
$\termone \gamma = \gamma(x)  \in \redgh{\Theta }{X}{H}$: we can see that it is indeed the case just by looking at the definition of $\redgr{\Theta}{X}{\domain \theta}{\Gamma} $ .
    \item  We consider the case where $\termone = \varone$, and $\varone$ is a real-type variable--either in $\Theta$ or in $\Gamma$. It means that the derivation of the typing judgment is:
      {\scriptsize
      $$ \AxiomC{$ \models \theta \Rightarrow \theta'$ }
 \LeftLabel{var-F}
\UnaryInfC{$ \Gamma', \Theta', \varone: \{ \alpha \in \typereal\}  {\imprr \theta {\theta'}} {\varone: \{ \alpha \in \typereal\}}$}
\DisplayProof $$}
 We suppose $X \subseteq \RR^n$, and $\gamma$ a substitution such that $\gamma \in  \redgr{\Theta}{X}{\domain \theta}{\Gamma} $  Observe that there are two possible cases: either $(x:F) \in \Gamma$, and then $\Gamma'$ denotes $\Gamma$ minus $(x:F)$, and $\Theta' = \Theta$, or $(x:F) \in \Theta$, and in this case $\Gamma' = \Gamma$, and $\Theta'$ denotes $\Theta$ minus $(x:F)$.
      \begin{varitemize}
      \item if $(x:F) \in \Theta$, then $\termone \gamma = x$. We suppose that $(x:F)$ is the i-th element of the list $\Theta$. Since $\gamma \in \redgr{\Theta}{X}{\domain \theta}{\Gamma} $, it holds that for every $a \in \RR$, $a \in \pi_i(X)$, implies that also $a \in \pi_i(\domain \theta)$. From there--and using the additional fact that $\models \theta \Rightarrow \theta'$, we can conclude that indeed $x \in \redgr{\Theta }{X}{\theta'}{F}$.
        \item if $(x:F) \in \Gamma$, then $\termone \gamma = \gamma(x)$. We suppose that $(x:F)$ is the $i$-th element of the ground part of $\Gamma$. We need to show that $\gamma(x) \in \redgr{\Theta}{X}{\theta'}{F}$, i.e.~that the function $\sem{\forget{\Theta} \imp \gamma(x):\typereal}$ sends continuously $X$ into $\domain{\theta'}$. Observe that this function coincides with $\pi_i \circ \sem \gamma$. Since $\gamma \in  \redgr{\Theta}{X}{\domain \theta}{\Gamma} $, we know that $\sem \gamma$ sends continuously $X$ into $\domain{\theta}$: we can conclude from here using the fact that $\models \theta \Rightarrow \theta'$. 
        \end{varitemize}
    \item We suppose now that $\termone =\abs{(\varone_1, \ldots, \varone_n)} \termtwo $. It means that the derivation of the typing judgment is as follows:
      {\scriptsize
        $$\AxiomC{$ \Gamma, \Theta, \varone_1: T_1, \ldots, \varone_n:T_n {\imprh {\psi'(\phi)}} \termtwo: T$}
\AxiomC{$\models \psi \wedge \theta \Rightarrow \psi'$}
\LeftLabel{abs}
\BinaryInfC{$ \Gamma, \Theta {\imprh {\theta}} \abs{(\varone_1, \ldots, \varone_n)} \termtwo : (T_1, \ldots, T_n) \tor{\psi(\phi)} T$}
\DisplayProof
$$}
      By applying the induction hypothesis, we obtain that:
\begin{equation}\label{eq:ind_hyp_abs_proof}
  \termtwo \in  \redgh{\Gamma'; \Theta}{\domain {\psi'}(\phi)}{ T} \quad \text{ with } \Gamma' = \Gamma,  \varone_1: T_1, \ldots, \varone_n:T_n .
  \end{equation}
      We suppose $X \subseteq \RR^n$, and $\gamma$ a substitution such that $\gamma \in  \redgr{\Theta}{X}{\domain \theta}{\Gamma} $. Our goal is to show that:
\begin{equation}\label{eq:goal_abs_proof}
\termone \gamma = \abs{(\varone_1, \ldots, \varone_n)} (\termtwo \gamma)  \quad \in \redgh{{\Theta }}{X(\eta)}{(T_1, \ldots, T_n) \tor{\psi(\phi)} T }\} .
\end{equation}
To show~\eqref{eq:goal_abs_proof}, we need to consider two families $u_i$, $\psi_i$ such that $\models \iota_1 \wedge \ldots \wedge \iota_n \Rightarrow \psi$,  and $u_i \in \redgh{\Theta}{Y^i(\iota_i)}{T_i}$, and to show that $(\termone \gamma)(u_1, \ldots,u_n) \in \redgh{\Theta}{X\cap_{1 \leq i \leq n} Y_i(\phi)}{T} $. (observe that $\iota_i$ is actually only defined when $T_i$ is a ground type: we denote $j_1, \ldots, j_l$ those indexes. We note $\iota = \iota_{j_1} \wedge \ldots \iota_{j_l}$. ).
Starting from $\gamma$, we build a substitution $\gamma':\vars \Gamma \cup \{\varone_1, \ldots,\varone_n \} \to \Lambda$ by taking:
$$\gamma'(x) := \gamma(x) \text{ when } x \in \vars \Gamma, u_i \text{ when } x = x_i.$$
We can see that $\gamma' \in \redgr{\Theta}{X\cap_{1 \leq i \leq n} Y_i}{ \domain{\theta \wedge \iota}}{\Gamma'}$. Observe that--since $\models \iota \Rightarrow \psi$ and $\models \theta \wedge \psi \Rightarrow \psi'$, it holds that $\models \iota \wedge \theta \Rightarrow \psi'$. As a consequence, --and using Lemma~\ref{lemma:monotony_logical_predicates}--it also holds that:
$$\gamma' \in \redgr{\Theta}{X\cap_{1 \leq i \leq n} Y_i}{ \domain{\psi'}}{\Gamma'}.$$
From there, we can deduce from~\eqref{eq:ind_hyp_abs_proof} that $\termtwo \gamma' \in \redgh{\Theta}{{(X\cap_{1 \leq i \leq n} Y_i)}(\phi)}{T}$, which concludes the proof.
    \item We look at the case where $\termone = \termfour (\termtwo_1, \ldots, \termtwo_m, \termthree_1, \ldots, \termthree_m)$. Looking at the typing rule for the application construct, we see that the derivation of the typing judgment is of the form:
{\scriptsize
      $$
\AxiomC{$\begin{array}{l}
  (\Gamma, \Theta {\imprh \theta} \termtwo_i: H_i)_{1 \leq i \leq m}
    \\ \Gamma, \Theta   {\imprh \theta} \termfour:  (H_1, \ldots, H_m, F_1, \ldots,F_n) \tor {\phi(\eta)} T
  \end{array}$}
\AxiomC{$
  \begin{array}{l}
    \models \phi_1 \wedge \ldots \wedge \phi_n \Rightarrow \phi \\
   (\Gamma, \Theta {\imprr \theta {\phi_j}} \termthree_j:F_j)_{1 \leq j \leq m}
   \end{array}
  $}
\LeftLabel{app}
\BinaryInfC{$ \Gamma, \Theta {\imprh {\theta(\eta)}}  \termfour (\termtwo_1, \ldots, \termtwo_m, \termthree_1, \ldots, \termthree_m) : T $}
\DisplayProof
$$}
By applying the induction hypothesis, we obtain that:
\begin{align}
 & \termfour \in \redgh{\Gamma; \Theta}{\domain \theta}{ (H_1, \ldots, H_m, F_1, \ldots,F_n) \tor {\phi(\eta)} T}. \label{eq:proof_app_func} \\
  & \termthree_j \in \redgr{\Gamma; \Theta}{\domain{\theta}}{\phi_j}{F_j}, \qquad  \termtwo_i \in \redgh{\Gamma; \Theta}{\domain{\theta}}{H_i}. \label{eq:proof_app_arg}
\end{align}
We suppose $X \subseteq \RR^n$, and $\gamma$ a substitution such that $\gamma \in  \redgr{\Theta}{X}{\domain \theta}{\Gamma} $. Our goal is to show that:
\begin{equation}\label{eq:goal_app_proof}
\termone \gamma = {\termfour \gamma}(\termtwo_1 \gamma, \ldots, \termtwo_m \gamma, \termthree_1 \gamma, \ldots, \termthree_m \gamma)  \quad \in \redgh{{\Theta }}{X(\eta)}{T}\} .
\end{equation}
Since $\gamma \in  \redgr{\Theta}{X}{\domain \theta}{\Gamma} $, we can deduce from~\eqref{eq:proof_app_func} and~\eqref{eq:proof_app_arg} that:
\begin{align*}
 & \termfour \gamma \in \redgh{\Theta}{X}{ (H_1, \ldots, H_m, F_1, \ldots,F_n) \tor {\phi(\eta)} T}.  \\
  & \termthree_j \gamma \in \redgr{\Theta}{X}{\phi_j}{F_j}; \qquad  \termtwo_i \gamma \in \redgh{\Theta}{X}{H_i}. 
\end{align*}
Looking at the definition of our logical predicate, we can conclude from here that~\eqref{eq:goal_app_proof} holds, which concludes the proof. 
  \item We look at the case where $\termone = \ifth{\termtwo}{\termthree}{\termfour}$. 

    We unfold $\Gamma$ and $\Theta$ as $\Gamma = \hpart \Gamma,(\varone_i: \{\alpha_i \in \typereal\})_{1 \leq i \leq n} $, and $\Theta =  (\vartwo_j: \{\beta_j \in \typereal\})_{1 \leq j \leq m}$ respectively.
    Looking at the typing rule for the conditional construct, we see that the derivation tree of the typing judgment of $\termone$ is of the form:
{\scriptsize
    \begin{center}

$$
\AxiomC{$
  \begin{array}{c}
      \Gamma, \Theta {\imprr {\theta_{\termtwo}}{ (\beta = 0 \vee \beta = 1)}} \termtwo: \{ \beta \in \typereal\}\\
       \Gamma, \Theta {\imprr {\theta_{(\termtwo,0)}}{(\beta = 0)}} \termtwo: \{ \beta \in \typereal \}\\
        \Gamma, \Theta {\imprr {\theta_{(\termtwo,1)}}{(\beta = 1)}} \termtwo: \{ \beta \in \typereal\}
      \end{array}$
}
\AxiomC{$\Gamma, \Theta {\imprh{\theta_{\termthree}(\eta)}} \termtwo:  T $}
\AxiomC{ $\Gamma, \Theta {\imprh{\theta_{\termfour}(\eta)}} \termthree:  T $ }
\AxiomC{~\eqref{side_cond_if_rule2},~\eqref{side_cond_if_rule3} }
\RightLabel{If}
\QuaternaryInfC{$ \Gamma, \Theta {\imprh {\theta(\eta)}} \ifth \termtwo \termthree \termfour : T $}
\DisplayProof
$$
 \end{center}
    The side-conditions~\eqref{side_cond_if_rule2}, ~\eqref{side_cond_if_rule3} are given as:
    \begin{varenumerate}
      \item\label{side_cond_if_rule2}  $\models \theta \Rightarrow  \left((\theta^{\termthree} \vee \theta^{\termfour}) \wedge (\theta^{(\termtwo,0)} \vee \theta^{\termfour}) \wedge (\theta^{(\termtwo,1)} \vee \theta^{\termthree}) \wedge (\theta_\termtwo \vee (\theta_\termthree \wedge \theta_\termfour))\right)$.
      \item \label{side_cond_if_rule3} 
        $ \forall (a_1, \ldots,a_{n}, b_1, \ldots, b_m) \in \RR^{n+m}$, $(a_1 = \alpha_1, \ldots, a_n = \alpha_n, b_1 = \beta_1, \ldots, b_m = \beta_m )\models \theta \wedge \neg \theta_{\termtwo}$ implies:  
  $$ \hpart \Gamma \imp \subst{\subst \termthree {x_1}{a_1} \ldots}{y_m}{b_m} \equiv^{ctx} \subst{\subst \termfour {x_1}{a_1} \ldots}{y_m}{b_m}.$$
 \end{varenumerate}
}
By appying the induction hypothesis, we obtain that:
\begin{align*}
  &\termthree \in \redgh{\Gamma; \Theta}{\domain{\theta_{\termthree}}(\eta)}{T}, \quad \termfour \in \redgh{\Gamma; \Theta}{\domain{\theta_{\termfour}}(\eta)}{T} \\
  & \termtwo \in \redgr{\Gamma; \Theta}{\domain{\theta_\termtwo}}{(\beta = 0 \vee \beta = 1)}{ \{\beta \in \typereal\}}  \\
  & \termtwo \in \redgr{\Gamma; \Theta}{\domain{\theta_{\termtwo,0}}}{(\beta = 0)}{ \{\beta \in \typereal\}} \cap \redgr{\Gamma; \Theta}{\domain{\theta_{\termtwo,1}}}{(\beta = 1)}{ \{\beta \in \typereal\}}
\end{align*}
  We suppose $X \subseteq \RR^n$, and $\gamma$ a substitution such that $\gamma \in  \redgr{\Theta}{X}{\domain \theta}{\Gamma} $. Our goal is to show that:
\begin{equation}
\termone \gamma = \ifth{\termtwo \gamma}{\termthree \gamma}{\termfour \gamma} \quad \in \redgh{{\Theta }}{X(\eta)}{T}\} .
\end{equation}
Looking at the definition of our predicate for substitutions, we see that our hypothesis $\gamma \in  \redgr{\Theta}{X}{\domain \theta}{\Gamma} $ also implies that  also $\gamma \in \redgr{\Theta}{ X_w}{\domain {\theta_w}}{\Gamma}$ for $w \in \{\termthree, \termfour, \termtwo, (\termtwo,1), (\termtwo,0)\}$, and where the $X_w$ are defined as:
$$  X_w = \sem{\gamma}^{-1}(\domain{\theta_w}) \cap X.$$
From there--and looking at how we define the open logical predicates in Definition~\ref{ldef:adm_substs_ref_types}--we can deduce that:
\begin{align*}
  &\termthree \gamma \in \redgh{\Theta}{X_{\termthree}(\eta)}{T}, \quad  \termfour \gamma \in \redgh{\Theta}{ X_{\termfour}(\eta)}{T} \\
  & \termtwo \gamma \in \redgr{ \Theta}{ X_{\termtwo}}{(\beta = 0 \vee \beta = 1)}{ \{\beta \in \typereal\}}  \\
  & \termtwo \gamma \in \redgr{\Theta}{X_{(\termtwo,0)}}{(\beta = 0)}{ \{\beta \in \typereal\}} \cap \redgr{\Theta}{X_{(\termtwo,1)}}{(\beta = 1)}{ \{\beta \in \typereal\}}
    \end{align*}
Observe that at this point, it is enough to show that we are allowed to apply Lemma~\ref{lemma:if-then-else_reduction}.  To do this, we need to check conditions~\ref{cond2:if-then-else_reduction}, \ref{cond3:if-then-else_reduction}.
  \begin{varitemize}
\item We first show Condition ~\ref{cond2:if-then-else_reduction} of Lemma~\ref{lemma:if-then-else_reduction}. First, we can see that our hypothesis~\ref{side_cond_if_rule2} implies:
\begin{align*}
  \domain \theta \subseteq & (\domain{\theta^{\termthree}} \cup \domain{\theta^{\termfour}}) \cap (\domain{\theta^{(\termtwo,0)}} \cup \domain{ \theta^{\termfour}})  \\
  & \cap (\domain{\theta^{(\termtwo,1)}} \cup \domain{\theta^{\termthree}}) \cap (\domain{\theta_\termtwo} \cup (\domain{\theta_\termthree} \cap \domain{\theta_\termfour})) .
\end{align*}
From there--and applying $\sem{\gamma^{-1}}$ to both side, we can deduce:
$$X \cap {\sem{\gamma}^{-1}}(\domain \theta) \subseteq  X^{\termthree} \cup X^{\termfour} \cap (X^{(\termtwo,0)} \cup X^{\termfour})  \cap (X^{(\termtwo,1)} \cup X^{\termthree}) \cap (X_\termtwo \cup (X_{\termthree} \cap X_\termfour)) $$
Since by hypothesis we know that $\gamma \in  \redgr{\Theta}{X}{\domain \theta}{\Gamma} $, it holds that $X \subseteq {\sem{\gamma}^{-1}}(\domain \theta)$ thus we can conclude that Condition~\ref{cond2:if-then-else_reduction} of Lemma~\ref{lemma:if-then-else_reduction} holds.
\item We now show Condition ~\ref{cond3:if-then-else_reduction} of Lemma~\ref{lemma:if-then-else_reduction}.  Let be $ a = (a_1, \ldots,a_n) \in X$, $a \in X \setminus X_{\termtwo}$. Looking at the way we define $X_\termtwo$--and since $\sem \gamma (X) \subseteq \domain \theta$--we can see that $\sem{\gamma}(a) \in \domain{\theta} \setminus {\domain{\theta_{\termtwo}}}$. We denote $\sem{\gamma}(a) = (a_1, \ldots, a_n, b_1, \ldots, b_{m})$. By our hypothesis~\ref{side_cond_if_rule3} we obtain that:
  $$ \hpart \Gamma \imp \subst{\subst \termthree {x_1}{a_1} \ldots}{y_m}{b_m} \equiv^{ctx} \subst{\subst \termfour {x_1}{a_1} \ldots}{y_m}{b_m}.$$
 
  Using the fact that--by definition of the denotational semantics--$\imp \gamma(y_j) \equiv^{ctx} b_j$, we can see that--by decomposing $\gamma$ into its higher-order part $\gamma_1$, and its ground part $\gamma_0$:
  $$ \hpart \Gamma \imp \subst{\subst {\termthree \gamma_0} {x_1}{a_1} \ldots}{x_n}{a_n} \equiv^{ctx} \subst{\subst {\termfour \gamma_0} {x_1}{a_1} \ldots}{x_n}{a_n}.$$
  At this point, we can finally conclude--by applying the substutution $\gamma_2$ to the higher-order variables in $\hpart \Gamma$ that:
  $$  \subst{\subst {\termthree \gamma_0 \gamma_1} {x_1}{a_1} \ldots}{x_n}{a_n} \equiv^{ctx} \subst{\subst {\termfour \gamma_0 \gamma_1} {x_1}{a_1} \ldots}{x_n}{a_n}.$$
  Observe that it is exactly Condition~\ref{cond3:if-then-else_reduction} of Lemma~\ref{lemma:if-then-else_reduction}, and it concludes the proof.
  \end{varitemize}
\end{varitemize}
 \end{proof}
        }
     
  From there, we can finally prove the main result of this section, i.e.~Theorem~\ref{theorem:correctness_refinement_types}, that states the correctness of our refinement type system. Indeed the fundamental lemma Lemma~\ref{lemma:fundamental_lemma_ref_types} above induces Proposition~\ref{proposition:well-typed-programs_in_pred_reftypes} as a corollary. Recall that Proposition~\ref{proposition:well-typed-programs_in_pred_reftypes} states that all typable programs are in the corresponding logical predicates: from there it is enough to look at the definition of the logical predicate for first-order programs to finally show the correctness of our type system. 

\section{Related Work}
\label{sect:related-work}

Logical relations are certainly one of the most well-studied concepts
in higher-order programming language theory. In their unary version,
they have been introduced by Tait~\cite{Tait67}, and further exploited
by Girard~\cite{Girard71} and Tait~\cite{Tait75} himself in giving
strong normalization proofs for second-order type
systems. The relational counterpart of realizability, namely logical
relations proper, have been introduced by Plotkin~\cite{plotkin1973lambda}, and further
developed along many different axes, and in particular towards calculi
with fixpoint constructs or recursive
types~\cite{DBLP:journals/toplas/AppelM01,AppelMellies,DBLP:conf/esop/Ahmed06}, 
probabilistic
choice~\cite{DBLP:conf/fossacs/BizjakB15}, or monadic and algebraic
effects~\cite{Goubault,DBLP:journals/pacmpl/BiernackiPPS18,Goubault}. 
Without any hope to be comprehensive, we may
refer to Mitchell's textbook on programming language theory for a
comprehensive account about the earlier, 
classic definitions~\cite{DBLP:books/daglib/0085577},
or to aforementioned papers for more recent developments.

Extensions of logical relations to open terms 
have been introduced by several 
authors
\cite{DBLP:conf/tlca/JungT93,DBLP:conf/mfcs/PittsS93,DBLP:conf/ppdp/Fiore02,DBLP:conf/lics/StatonYWHK16,DBLP:conf/icfp/BowmanA15} 
and
were explicitly referred to as \emph{open logical relations} 
in \cite{DBLP:conf/aplas/ZhaoZZ10}. 
However, to the best of authors' knowledge, all the aforementioned works 
use open logical relations for specific purposes, and do not investigate 
their applicability as a general methodology.

Special cases of our Containment Theorem can be found in many papers,
typically as auxiliary results. As already mentioned, an example is
the one of higher-order polynomials, whose first-order terms are
proved to compute proper polynomials in many
ways~\cite{DBLP:journals/siamcomp/KapronC96,baillot2012higher}, none of them
in the style of logical relations. 
The Containment Theorem itself can be derived by a previous result by 
Lafont \cite{lafont1988logiques} (see also Theorem 4.10.7 in 
\cite{DBLP:books/daglib/0031706}). Contrary to such a result, however, 
our proof of the Containment Theorem is entirely syntactical and consists 
of a straightforward application of differential logical relations.  

Algorithms for automatic
differentiation have recently been extended to higher-order
programming 
languages~\cite{Efficient/AD/jones,DBLP:journals/toplas/PearlmutterS08,DBLP:journals/lisp/SiskindP08,DBLP:journals/jfp/ManzyukPRRS19,DBLP:conf/popl/PearlmutterS07}, and have been 
investigated from a semantical perspective in \cite{DBLP:journals/pacmpl/BrunelMP20,DBLP:journals/pacmpl/AbadiP20} relying on insights from linear logic and denotational 
semantics. 
Continuity and robustness analysis of imperative first-order
programs by way of program logics is the topic of study of a series
of papers by Chaudhuri and co-authors~\cite{DBLP:journals/cacm/ChaudhuriGL12,ChaudhuriGL10,DBLP:conf/sigsoft/ChaudhuriGLN11}. None
of them, however, deal with higher-order programs.
\section{Conclusion and Future Work}
We have showed how a mild variation on the concept of a logical
relation can be fruitfully used for proving both predicative and
relational properties of higher-order programming languages, when
such properties have a first-order, rather than a ground ``flavor''.
As such, the added value of this contribution is not much in
the technique itself, but in showing how it is extremely
useful in heterogeneous contexts, this way witnessing the
versatility of logical relations.

The three case studies, and in particular the correctness of automatic
differentiation and refinement type-based continuity analysis, are
given as proof-of-concepts, but this does not mean they do not
deserve to be studied more in depth. An example of an interesting
direction for future work is the extension of our correctness proof
from Section~\ref{sect:automatic-differentiation} to
backward propagation differentiation algorithms. Another one consists
in adapting the refinement type system of Section~\ref{sec:cont} to
deal with differentiability.  That would of course require a
substantial change in the typing rule for conditionals, which should
take care of checking not only continuity, but also differentiability
at the critical points. It would also be interesting to implement the
refinement type system using standard SMT-based approaches.

Finally, the authors plan to investigate extensions of open logical relations 
to non-normalizing calculi, as well as to non-simply typed calculi 
(such as calculi with polymorphic or recursive types).
\bibliographystyle{splncs04}
\bibliography{main}

\end{document}

%% file: macros.tex
\newtheorem{notation}{Notation}
\newenvironment{varitemize}
{
\begin{list}{\labelitemi}
{\setlength{\itemsep}{0pt}
 \setlength{\topsep}{0pt}
 \setlength{\parsep}{0pt}
 \setlength{\partopsep}{0pt}
 \setlength{\leftmargin}{15pt}
 \setlength{\rightmargin}{0pt}
 \setlength{\itemindent}{0pt}
 \setlength{\labelsep}{5pt}
 \setlength{\labelwidth}{10pt}
}}
{
 \end{list}
}
\newcounter{numberone}
\newenvironment{varenumerate}
{
\begin{list}{\arabic{numberone}.}
{
  \usecounter{numberone}
  \setlength{\itemsep}{0pt}
  \setlength{\topsep}{0pt}
  \setlength{\parsep}{0pt}
  \setlength{\partopsep}{0pt}
  \setlength{\leftmargin}{15pt}
  \setlength{\rightmargin}{0pt}
  \setlength{\itemindent}{0pt}
  \setlength{\labelsep}{5pt}
  \setlength{\labelwidth}{15pt}
}}
{
\end{list}
} 
\newcommand{\STLambda}{\Lambda^{\times, \to}}
\newcommand{\STLambdareal}{\Lambda^{\times, \to, \typereal}}
\newcommand{\STLambdarealfun}{\Lambda^{\times, \to, \typereal}_{\funsymbols}}
\newcommand{\STLambdarealcont}{\Lambda^{\times, \to, \typereal}_{\contfun}}
\newcommand{\STLambdarealderiv}{\Lambda^{\times, \to, \typereal}_{\derivfun}}

\newcommand{\hide}[1]{}
\RequirePackage{ifthen}
\newcommand{\typeof}{0}
\newcommand{\longv}[1]{\ifthenelse{\equal{\typeof}{0}}{}{#1}}
\newcommand{\shortv}[1]{\ifthenelse{\equal{\typeof}{0}}{#1}{}}

\newcommand{\tor}[1]{ \stackrel{#1}{\to}}
\newcommand{\torr}[2]{{ \stackrel{#1 \leadsto #2}{\to}}}
\newcommand{\variables}{\mathcal{V}}

\newcommand{\lan}{\langle}
\newcommand{\ran}{\rangle}

\newcommand{\forget}[1]{\overline{{#1}}}
\newcommand{\domain}[1]{\text{Dom}(#1)}

\newcommand{\varone}{x}
\newcommand{\vartwo}{y}
\newcommand{\vars}[1]{\text{Vars}(#1)}
\newcommand{\abs}[2]{\lambda #1.#2}
\newcommand{\ifth}[3]{\mathtt{if}\ #1\ \mathtt{then}\ #2\ \mathtt{else}\ #3 }

\newcommand{\termone}{t}
\newcommand{\termtwo}{s}
\newcommand{\termthree}{p}
\newcommand{\termfour}{q}
\newcommand{\typeone}{\tau}
\newcommand{\typetwo}{\sigma}

\newcommand{\funsymbol}{f}
\newcommand{\funsymbols}{\mathfrak{F}}
\newcommand{\funpred}{\mathcal{F}}

\newcommand{\contfun}{\mathfrak{C}}
\newcommand{\contpred}{\mathcal{C}}
\newcommand{\derivfun}{\mathfrak{D}}

\newcommand{\makereal}[1]{\underline{#1}}
\newcommand{\RR}{\mathbb{R}}
\newcommand{\NN}{\mathbb N}
\newcommand{\typereal}{\mathtt{R}}

\newcommand{\comp}{\circ}

\newcommand{\imp}{\vdash}
\newcommand{\impr}{\, \vdash_{\mathtt{r}} \,}
\newcommand{\imprr}[2]{\, \stackrel{#1 \leadsto #2}{\vdash_{\mathtt{r}}} \,}
\newcommand{\imprh}[1]{\, \stackrel{#1}{\vdash_{\mathtt{r}}} \,}
\newcommand{\exprs}{\mathcal E}
\newcommand{\varlone}{\alpha}

\newcommand{\hh}{\hdots}
\newcommand{\cc}{\cdots}
\newcommand{\bnf}{\;::=\;}
\newcommand{\midd}{\; \; \mbox{\Large{$\mid$}}\;\;} 
\newcommand{\sem}[1]{\llbracket #1 \rrbracket}

\newcommand{\subst}[3]{#1[#3/#2]}
\newcommand{\support}[1]{\mathsf{supp}(#1)}

\newcommand{\redgr}[4]{\contpred (#1, {#2 \leadsto #3},#4)}
\newcommand{\redgh}[3]{\contpred(#1, #2,#3)}

\newcommand{\gpart}[1]{\,G #1\,}
\newcommand{\hpart}[1]{\,H #1 \,}

\newcommand{\envone}{\Gamma}
\newcommand{\fst}[1]{#1.1} 
\newcommand{\snd}[1]{#1.2}
\newcommand{\ith}[1]{#1.i}
\newcommand{\envreal}{\Theta}
\newcommand{\substone}{\gamma}
\newcommand{\substtwo}{\delta}

\newcommand{\de}[1]{\mathtt{D} #1}
\newcommand{\devar}[1]{\mathtt{d} #1}

\newcommand{\dual}[2]{\mathtt{dual}_{#1}(#2)}
\newcommand{\deriv}[2]{\mathtt{deriv}(#1,#2)}

\newcommand{\relone}{\mathbin{\mathcal{R}}}

\newtheorem*{mytheorem}{Theorem 1}

\newcommand{\ctheorem}{Containment Theorem}